%% file: main.tex
\documentclass[sigplan,10pt]{acmart}
\makeatletter
\@ifundefined{@ACM@copyright@check@cc}{}{%
  \let\@ACM@copyright@check@cc\@empty
}
\makeatother
\AtBeginDocument{\settopmatter{printfolios=false}}

\copyrightyear{2027}\acmYear{2027}
\setcopyright{cc}
\setcctype{by}
\acmConference[EuroSys '27]{22nd European Conference on Computer Systems}{April 19-23, 2027}{Rabat, Morocco}
\acmBooktitle{22nd European Conference on Computer Systems (EuroSys '27), April 19-23, 2027, Rabat, Morocco}
\acmDOI{10.1145/3842654.3848563}
\acmISBN{979-8-4007-2971-3/2027/04}



\input{80-authors}

\input{90-packages}
\input{91-commands}

\begin{document}
\title{\MANTA: Machine Learning Augmented Tiering Advisor}

\begin{abstract}
\input{00-abstract}
\end{abstract}

\maketitle %

\section{Introduction}
\label{sec:introduction}
\input{10-introduction}

\section{Background}
\label{sec:background}
\input{20-background}

\section{Characterizing the Limits of Heuristic Tiering Policies}
\label{sec:motivation}
\input{30-motivation}

\section{Design}
\label{sec:design}
\input{40-design}

\section{Evaluation}
\label{sec:evaluation}
\input{50-evaluation}

\section{Conclusion}
\label{sec:conclusion}
\input{70-conclusion}

\section*{Acknowledgements}
\input{71-acknowledgements}

\bibliographystyle{ACM-Reference-Format}
\bibliography{references}

\end{document}

%% file: 80-authors.tex
\author{Johannes Freischuetz}
\email{freischuetz@cs.wisc.edu}
\orcid{0009-0004-2667-4852}
\affiliation{%
  \institution{University of Wisconsin -- Madison}
  \city{Madison}
  \state{Wisconsin}
  \country{USA}
  \postcode{53706}
}

\author{Kiet Pham}
\email{kvpham@wisc.edu}
\orcid{0009-0007-2641-8228}
\affiliation{%
  \institution{University of Wisconsin -- Madison}
  \city{Madison}
  \state{Wisconsin}
  \country{USA}
  \postcode{53706}
}

\author{Sujay Yadalam}
\email{sujay@cs.utexas.edu}
\orcid{}
\affiliation{%
  \institution{University of Texas at Austin}
  \city{Austin}
  \state{Texas}
  \country{USA}
}

\author{Konstantinos Kanellis}
\authornote{This work was conducted while at the University of Wisconsin -- Madison.}
\email{kkanelli@amazon.com}
\orcid{0009-0006-3776-3863}
\affiliation{%
  \institution{Amazon Web Services}
  \city{New York}
  \state{New York}
  \country{USA}
}

\author{Michael Swift}
\email{swift@cs.wisc.edu}
\orcid{}
\affiliation{%
  \institution{University of Wisconsin -- Madison}
  \city{Madison}
  \state{Wisconsin}
  \country{USA}
  \postcode{53706}
}

\author{Shivaram Venkataraman}
\email{shivaram@cs.wisc.edu}
\orcid{0000-0001-9575-7935}
\affiliation{%
  \institution{University of Wisconsin -- Madison, ETH Zurich}
  \city{Zurich}
  \country{Switzerland}
}

%% file: 90-packages.tex
\usepackage{color,soul}
\usepackage[dvipsnames]{xcolor}
\usepackage{hyperref}
\usepackage{enumitem}
\usepackage{subcaption}
\usepackage{amsmath}
\usepackage[noend]{algpseudocode}
\usepackage{algorithm}
\usepackage[]{mdframed}
\usepackage[utf8]{inputenc}   %
\usepackage[T1]{fontenc}      %
\usepackage{hyperref}         %
\usepackage{url}              %
\usepackage{booktabs}         %
\usepackage{array}            %
\usepackage{nicefrac}         %
\usepackage{amsfonts}         %
\usepackage{microtype}        %
\usepackage{graphicx}         %
\usepackage{sidecap, caption} %
\usepackage{multirow}         %
\usepackage{enumitem}         %
\usepackage{mathtools}
\usepackage{dsfont}
\usepackage{xspace}
\usepackage{tikz}
\usepackage{pifont}%

\usepackage[toc,page]{appendix}
\usepackage{fancyhdr}

%% file: 91-commands.tex
\newcommand{\plotlegend}[2]{%
  \makebox[\linewidth][c]{\hspace*{\dimexpr#1+#1\relax}#2}\par
}

\newcommand{\eat}[1]{}

\newif\ifshowchanges
\showchangestrue
\ifshowchanges

\else

\fi

\algrenewcommand\algorithmicrequire{\textbf{Input:}}
\algrenewcommand\algorithmicensure{\textbf{Output:}}
\algnewcommand\algorithmicforeach{\textbf{for each}}
\algdef{S}[FOR]{ForEach}[1]{\algorithmicforeach\ #1\ \algorithmicdo}

\newcommand{\cmark}{\color{green}{\ding{51}}}%
\newcommand{\xmark}{\color{red}{\ding{55}}}%

\DeclareRobustCommand{\figcallout}[1]{\ding{\numexpr181+#1\relax}}

\makeatletter
\newcommand\bigDiamond{\mathop{\mathpalette\bigDi@mond\relax}}
\newcommand\bigDi@mond[2]{%
  \vcenter{\hbox{\m@th
    \scalebox{\ifx#1\displaystyle 2\else1.2\fi}{$#1\Diamond$}%
  }}%
}
\newcommand\bigLozenge{\mathop{\mathpalette\bigL@zenge\relax}}
\newcommand\bigL@zenge[2]{%
  \vcenter{\hbox{\m@th
    \scalebox{\ifx#1\displaystyle 2\else1.2\fi}{$#1\blacklozenge$}%
  }}%
}

\newcommand{\MANTA}{\texttt{MANTA}\xspace}

\newcommand{\ARMS}{\texttt{ARMS}\xspace}
\newcommand{\TPP}{\texttt{TPP}\xspace}
\newcommand{\nomad}{\texttt{NOMAD}\xspace}
\newcommand{\colloid}{\texttt{Colloid}\xspace}

\newcommand{\hybridtier}{\texttt{HybridTier}\xspace}
\newcommand{\hemem}{\texttt{HeMem}\xspace}
\newcommand{\memtis}{\texttt{Memtis}\xspace}
\newcommand{\kleio}{\texttt{Kleio}\xspace}
\newcommand{\coeus}{\texttt{Coeus}\xspace}
\newcommand{\cronus}{\texttt{Cronus}\xspace}
\newcommand{\cori}{\texttt{Cori}\xspace}
\newcommand{\IDT}{\texttt{IDT}\xspace}
\newcommand{\artmem}{\texttt{ArtMem}\xspace}
\newcommand{\learningtorank}{Learning-to-Rank\xspace}

\gdef\chomp{\texttt{ChOMP}\xspace}
\gdef\chopt{\texttt{CHOPT}\xspace}
\gdef\choptpp{\texttt{CHOPT++}\xspace}
\gdef\mira{\texttt{Mira}\xspace}
\gdef\vulcan{\texttt{Vulcan}\xspace}
\gdef\lleaves{\texttt{lleaves}\xspace}
\gdef\lightgbm{\texttt{LightGBM}\xspace}
\gdef\smac{\texttt{SMAC}\xspace}
\gdef\mess{\texttt{MESS}\xspace}
\gdef\gurobi{\texttt{Gurobi}\xspace}
\gdef\cloudlabnode{\texttt{c220g5}\xspace}
\gdef\xeongold{\texttt{Intel(R) Xeon(R) Gold 6226R}\xspace}
\gdef\hynixdram{\texttt{SK Hynix 32GB PC4-25600 DDR4-3200MHz}\xspace}
\gdef\optanepmem{\texttt{Intel Optane Persistent Memory 100 Series}\xspace}
\gdef\ewmatwo{\texttt{ewma\_2}\xspace}
\gdef\ewmafive{\texttt{ewma\_5}\xspace}
\gdef\ewmatwenty{\texttt{ewma\_20}\xspace}
\gdef\ewmahundred{\texttt{ewma\_100}\xspace}
\gdef\globalavgaccesses{\texttt{global\_avg\_accesses}\xspace}
\gdef\pageage{\texttt{age}\xspace}
\gdef\rratiotwenty{\texttt{r\_ratio\_20}\xspace}
\gdef\rratiohundred{\texttt{r\_ratio\_100}\xspace}
\gdef\groupabove{\texttt{group\_above}\xspace}
\gdef\groupbelow{\texttt{group\_below}\xspace}
\gdef\nearbypages{\texttt{nearby\_pages}\xspace}
\gdef\nearbyewmavariance{\texttt{nearby\_ewma5\_var}\xspace}
\gdef\pebsloadmiss{\texttt{MEM\_LOAD\_L3\_MISS\_RETIRED.LOCAL\_DRAM}\xspace}
\gdef\pebsstores{\texttt{MEM\_INST\_RETIRED.ALL\_STORES}\xspace}
\gdef\lthreehitsnoop{\texttt{L3\_HIT.ANY\_SNOOP}\xspace}
\gdef\chompfullname{\underline{Ch}unked \underline{O}ffline \underline{M}emory \underline{P}olicy\xspace}
\newcommand{\faiss}{\texttt{FAISS}\xspace}

\gdef\bc{\texttt{BC}\xspace}
\gdef\pr{\texttt{PR}\xspace}
\gdef\bckron{\texttt{BC-Kron}\xspace}
\gdef\bctwitter{\texttt{BC-Twitter}\xspace}
\gdef\bckronbig{\texttt{BC-Kron-80GB}\xspace}
\gdef\prkron{\texttt{PR-Kron}\xspace}
\gdef\prkronbig{\texttt{PR-Kron-80GB}\xspace}
\gdef\prtwitter{\texttt{PR-Twitter}\xspace}
\gdef\xsbench{\texttt{XSBench}\xspace}
\gdef\multigrid{\texttt{NAS MultiGrid}\xspace}
\gdef\duckdb{\texttt{DuckDB}\xspace}

%% file: 00-abstract.tex
Memory tiering has been used to expand memory capacity, particularly in datacenters, by combining fast DRAM with slower tiers, including CXL-attached memory.
Its effectiveness depends on keeping useful pages in the fast tier, but existing heuristic policies can lag behind changing hot sets in phased or bursty workloads.
To explore these limitations, we introduce \chomp, a scalable offline optimizer that minimizes placement and bandwidth-sensitive migration costs.
We then develop a trace-driven simulator that uses this reference to identify performance opportunities for online policies.
Motivated by these results, \MANTA predicts future page usefulness from runtime access features and integrates a lightweight learned model into \ARMS.
Across eight workloads on emulated CXL, \MANTA achieves geometric-mean speedups over \ARMS of 1.12$\times$ and 1.08$\times$ at 4~GB of fast memory on Linux 6.2 and 6.18, respectively; across six Optane workloads, it achieves 1.69$\times$.
On individual workloads, \MANTA is up to 1.25$\times$ faster than \ARMS with emulated CXL on Linux 6.2, 1.21$\times$ on Linux 6.18, and 5.6$\times$ with Optane.

%% file: 10-introduction.tex
Memory is an increasing fraction of datacenter cost~\cite{tmo,TPP}, motivating tiered deployments that combine fast local DRAM with larger, cheaper CXL or persistent memory~\cite{microsoft_cxl_deployment,cxl_pond_msft,alibaba_cloud_cxl_db,cxl_overview,guide_to_pmem}.
We refer to local DRAM as the \emph{fast} tier and CXL or persistent memory as the \emph{slow} tier.
Reported CXL latencies are often 200--300~ns versus roughly 100~ns for DRAM, with Optane slower still~\cite{MESS,berger_cxl_perf,optane_perf2}.
Effective tiering must keep useful pages in fast memory~\cite{hitting_memory_wall} while deciding when migration benefits outweigh copying costs and bandwidth contention.
A page must remain useful long enough for its fast-memory accesses to repay the cost of moving it.
Migration timing also matters because copying pages competes with the application for bandwidth.

Existing systems~\cite{SoarAlto,flexmem,thermostat,nimble_page_management,HeMem,TPP,ARMS,MEMTIS} use heuristics based on recency, frequency, page age, or bandwidth pressure~\cite{COLLOID}.
These policies are reactive, promoting or demoting pages when recent behavior crosses a threshold.
Access history often suffices for stable workloads~\cite{flexmem,nomad}, but phased or bursty workloads can change hot sets faster than policies adapt.
Slow reactions leave useful data in the slow tier or waste scarce fast-tier capacity on stale hot pages.

\input{figures/bc_kron_heatmap.tex}

Figure~\ref{fig:bc-kron-heatmap} illustrates the problem on Betweenness Centrality (\bc) over the Kron graph (\bckron).
\ARMS~\cite{ARMS} retains pages from the previous hot phase while currently useful pages remain in slow memory.
Much of the benefit is lost because the hot set changes again almost immediately after migration.

To systematically characterize these missed opportunities, we compare online policies with an offline reference that knows the entire access trace.
An offline reference can expose missed opportunities even when several online policies make similar placement mistakes.
Existing offline formulations~\cite{CHOPT} have two limitations: solving a 10-second trace with around a billion accesses takes more than 60 hours, and fixed migration penalties ignore contention with concurrent memory traffic~\cite{MESS}.
We develop \chomp, a Chunked Offline Memory Policy, that aggregates accesses within intervals and accounts for bandwidth-sensitive migration costs.
It minimizes migration costs minus fast-memory access savings among schedules that migrate at interval boundaries.
\chomp solves traces up to 630$\times$ longer than \chopt in a similar time.

The resulting schedule gives us concrete placements and migration times to compare with online policies.
To evaluate this offline schedule, we build a trace-driven simulator calibrated against measured runtimes (Section~\ref{sec:motivation:simulator-fidelity}).
For \bckron with 2~GB of fast memory, \MANTA reduces measured runtime from 241.3~s with \ARMS to 221.3~s, a 1.09$\times$ speedup (Figure~\ref{fig:sim_eval_bc-kron}).
Comparing these measurements with the simulated \chomp reference shows that \MANTA closes much of the gap while leaving room for further gains (Section~\ref{sec:eval:simulator-evaluation}).
The remaining gap exposes a timing problem: late promotions and demotions lose performance when a policy recognizes a page's usefulness only after its accesses occur (Section~\ref{sec:motivation:simulator}).

Recurring access patterns provide a way to anticipate that usefulness from information available at runtime.
\MANTA predicts a page's discounted future-access count, giving greater weight to nearer accesses (Equation~\ref{eq:discounted_value}).
Its model combines recent activity, longer-term history, and spatial context, learning how these signals interact for each application so it can identify useful pages before reactive hotness scores catch up.

We integrate these predictions into \ARMS's periodic tiering loop, retaining its access tracking, migration execution, and capacity management.
We target recurring, profileable applications, training application-specific models on prior runs.
\MANTA makes its predictions using a lightweight tree model compiled to native code, with features commonly used in prior work.
We introduce sample-based virtual time steps to make these features less sensitive to execution speed.
The compiled model scores more than 10,000 pages in under 20~ms.

We evaluate \MANTA across emulated CXL and Optane, eight workloads, and two Linux kernels, comparing with \ARMS, \hybridtier, \nomad, \TPP, and \memtis~\cite{ARMS,hybridtier,nomad,TPP,MEMTIS}.
At 4~GB of fast memory, geometric-mean speedups over \ARMS are 1.12$\times$ on Linux 6.2, 1.08$\times$ on Linux 6.18, and 1.69$\times$ across six Optane workloads.
Across evaluated workloads and capacities, maximum speedups over \ARMS are approximately 1.25$\times$ on Linux 6.2, 1.21$\times$ on Linux 6.18, and 5.6$\times$ on Optane.
On Linux 6.2, \MANTA has the highest geometric-mean speedup at every capacity, outperforming \memtis by 1.28$\times$ at 4~GB and 1.03--1.11$\times$ at larger capacities.

%% file: figures/bc_kron_heatmap.tex
\begin{figure}[t]
\centering

\begin{subfigure}[t]{0.47\linewidth}
    \centering
    \includegraphics[width=\linewidth,page=7]{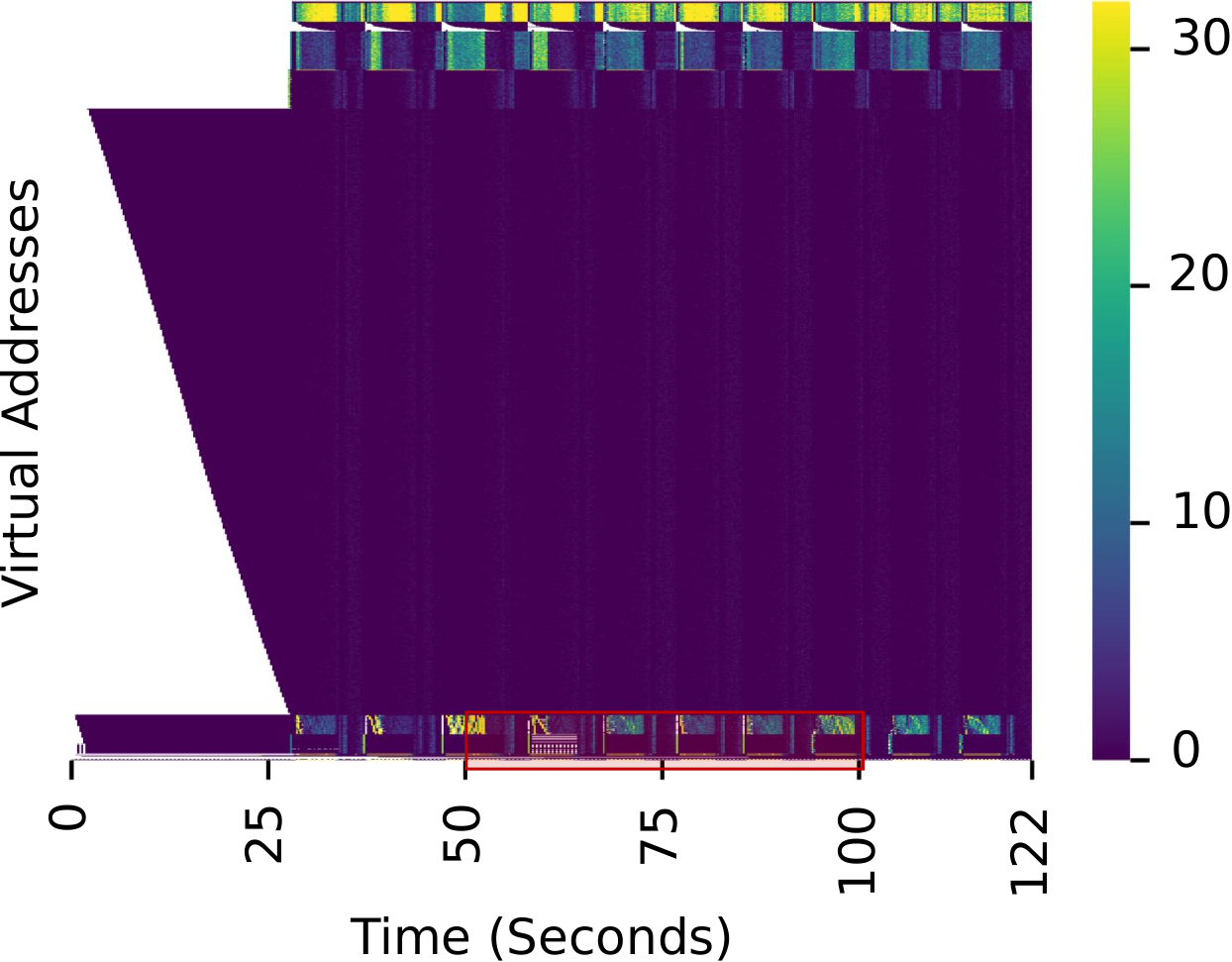}
    \caption{\bckron access heatmap. The highlighted interval is enlarged at right.}
    \label{fig:bc-kron-heatmap-full}
\end{subfigure}%
\hfill%
\begin{subfigure}[t]{0.47\linewidth}
    \centering
    \includegraphics[width=\linewidth,page=8]{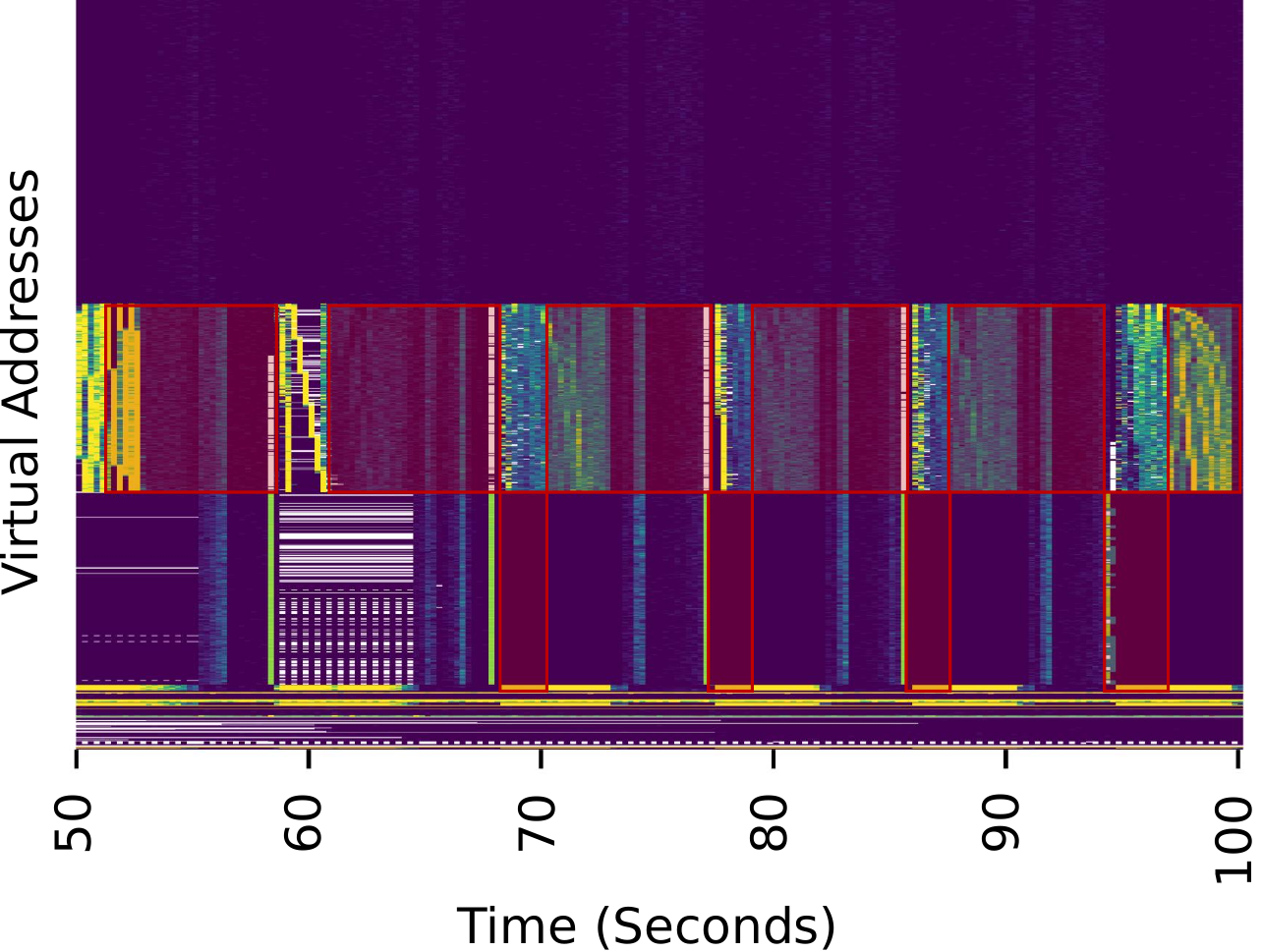}
    \caption{Approximate \ARMS placement with 4~GB of fast memory.}
    \label{fig:bc-kron-heatmap-dram}
\end{subfigure}%
\caption{Access intensity for \bckron on the left, with approximate page placements on the right.
Delayed migrations cause much of the benefit to be missed.}

\Description{Two side-by-side \bckron heatmaps.
The left panel shows the full page-access heatmap over time, and the right panel shows a zoomed-in view of the pages resident in DRAM during the highlighted interval.}
\label{fig:bc-kron-heatmap}
\end{figure}

%% file: 20-background.tex
Earlier work explored page placement and data tiering in heterogeneous memory systems~\cite{page_placement, data_tiering, thermostat, nimble_page_management}.
Online tiered-memory systems use observed page accesses to decide which pages to promote to fast memory and which to demote to slow memory.
Systems such as \TPP~\cite{TPP}, \ARMS~\cite{ARMS}, \hybridtier~\cite{hybridtier}, \nomad~\cite{nomad}, and \hemem~\cite{HeMem} combine low-overhead tracking with heuristic estimates of page hotness.
\colloid~\cite{COLLOID} also incorporates contention and latency measurements into placement decisions.
These policies use recent measured behavior to guide future placement, making hotness estimation a common component despite differences in tracking and migration mechanisms.
\mira~\cite{mira} takes a complementary approach through program-guided slow-memory management rather than transparent page tiering.

Tiered-memory policies differ in the signals they use, where their decisions execute, and the granularity at which they manage memory.
Table~\ref{tab:learned-tiering-comparison} summarizes these differences for the systems most closely related to \MANTA.

\input{figures/learned_tiering_comparison.tex}

Learning-based policies explore alternatives to heuristic hotness estimation.
\kleio, \coeus, and \cronus use learned access predictions or select pages and page groups for learned management~\cite{kleio, Coeus, Cronus}.
These works primarily explore prediction in offline or simulator-driven settings, with inference costs posing a challenge for an online policy deployment.
Other learning-based approaches act on different decisions: \IDT and \artmem use reinforcement learning to tune demotion policies or migration scope~\cite{idt, artmem}, while \learningtorank predicts an offline ranking of application objects for static placement without migration~\cite{learning_to_rank_objects}.

Table~\ref{tab:learned-tiering-comparison} contrasts direct prediction of future page usefulness with heuristic tuning and static object placement.
\MANTA combines direct page-level prediction with online execution and dynamic migration.
It predicts a lightweight discounted future-usefulness score that can replace the hotness estimate in an existing ranking-based policy, while the surrounding system continues to handle the migration mechanism.
This integration requires scoring pages quickly enough to fit within the existing periodic tiering loop.
Migration timing is a complementary policy decision: \cori~\cite{cori} uses application data-reuse information to tune the frequency of data movement across memory tiers.

To put these online placement and migration decisions in context, we consider policies that know the full access trace.
Belady's MIN is the classical offline benchmark for single-tier caching, giving the optimal eviction rule~\cite{Belady}.
Prior work develops practical bounds on optimal caching for variable-size objects~\cite{caching_optimal_bounds}.
Learning Relaxed Belady shows how oracle-style caching can be approximated in other domains such as CDN caches~\cite{LearnedBelady}.
For tiered memory, \chopt formulates offline-optimal placement across heterogeneous cache, memory, and storage tiers with asymmetric access costs, and uses sampling to approximate longer traces~\cite{CHOPT}.
Section~\ref{sec:motivation} examines the limitations of these approaches for interval-based page migration and motivates a reference policy that accounts for migration interference and bandwidth contention.

%% file: figures/learned_tiering_comparison.tex
\begin{table*}[t]
\definecolor{tablecheckgreen}{HTML}{4C9A67}
\definecolor{tablexred}{HTML}{E63325}
\renewcommand{\cmark}{\textcolor{tablecheckgreen}{\ding{51}}}
\renewcommand{\xmark}{\textcolor{tablexred}{\ding{55}}}
\centering
\footnotesize
\setlength{\tabcolsep}{3pt}
\renewcommand{\arraystretch}{1.0}
\begin{tabular}{@{}l @{\hspace{10pt}} l
  @{\hspace{10pt}} c
  @{\hspace{10pt}} c
  @{\hspace{10pt}} c
  @{\hspace{10pt}} c
  @{\hspace{10pt}} l@{}}
\toprule
\raisebox{1.2ex}{\textbf{System}} &
\raisebox{1.2ex}{\textbf{Decision/Output}} &
\shortstack{\textbf{ML-}\\\textbf{Based}} &
\shortstack{\textbf{Direct Hotness}\\\textbf{Prediction}} &
\shortstack{\textbf{Online}\\\textbf{System}} &
\shortstack{\textbf{Dynamic}\\\textbf{Migration}} &
\shortstack[l]{\textbf{Decision}\\\textbf{Granularity}} \\
\midrule
\TPP~\cite{TPP} & Recent access frequency and age & \xmark & \xmark & \cmark & \cmark & Pages \\
\memtis~\cite{MEMTIS} & Hotness class and page size & \xmark & \xmark & \cmark & \cmark & Pages \\
\nomad~\cite{nomad} & Recent page hotness & \xmark & \xmark & \cmark & \cmark & Pages \\
\hybridtier~\cite{hybridtier} & Access-frequency hotness & \xmark & \xmark & \cmark & \cmark & Pages \\
\ARMS~\cite{ARMS} & Adaptive page-hotness score & \xmark & \xmark & \cmark & \cmark & Pages \\
\midrule
\kleio~\cite{kleio} & Next-epoch page accesses & \cmark & \cmark & \xmark & \cmark & Selected Pages \\
\coeus~\cite{Coeus} & Page-group access patterns & \cmark & \cmark & \xmark & \cmark & Selected Page Groups \\
\cronus~\cite{Cronus} & Groups for learned page management & \cmark & \cmark & \xmark & \cmark & Selected Page Groups \\
\cori~\cite{cori} & Migration cadence & \xmark & \xmark & \xmark & \cmark & System-wide knob \\
\learningtorank~\cite{learning_to_rank_objects} & Object performance rank & \cmark & \xmark & \xmark & \xmark & Objects \\
\IDT~\cite{idt} & Demotion-policy parameters & \cmark & \xmark & \cmark & \cmark & Heuristic Knobs \\
\artmem~\cite{artmem} & Hotness threshold and migration scope & \cmark & \xmark & \cmark & \cmark & Heuristic Knobs \\
\midrule
\MANTA & Discounted future page usefulness & \cmark & \cmark & \cmark & \cmark & Pages \\
\bottomrule
\end{tabular}

\caption{Memory-tiering approaches. Direct hotness prediction estimates future page usefulness. Online policies run during application execution, as opposed to in a simulator. Dynamic migration moves pages between tiers online.}

\Description{Comparison of tiered-memory placement and migration approaches.
ML-based systems use machine learning to change their tiering decisions dynamically.
Systems that use ``direct hotness predictions'' forecast future access behavior or usefulness for pages or page groups.
Online systems are those which can be deployed in a real-time system.
Systems supporting dynamic migration can migrate pages rather than relying on static allocations.}
\label{tab:learned-tiering-comparison}
\end{table*}

%% file: 30-motivation.tex
A useful way to understand the limitations of online tiering policies is to compare them with an offline policy that has full knowledge of the trace.
Classical offline caching policies, such as Belady's MIN~\cite{Belady}, however, are insufficient for tiered memory because they ignore migration costs and bandwidth contention~\cite{COLLOID}.
A Belady-style policy can migrate too aggressively in a tiered-memory system~\cite{CHOPT}.
It also ignores the potential benefit of distributing accesses across tiers when one memory link approaches saturation.

\input{figures/CHOPT.tex}

The state-of-the-art offline policy for tiered memory is \chopt~\cite{CHOPT}; Figure~\ref{fig:CHOPT} shows its access-level graph with our time-step grouping overlaid.
\chopt formulates placement as a min-cost-flow problem in which each unit of flow represents one unit of fast-memory capacity.
A flow through the graph selects which pages occupy fast memory over the access trace, while edge costs encode fast-memory benefits and migration penalties.
Here, an optimal policy is a placement and migration schedule that minimizes total modeled cost over the trace among schedules that satisfy the capacity and migration constraints in \chopt's formulation.

This formulation is powerful because it can recover the optimal migration decisions for a trace under its cost model.
It has two limitations for our setting.
First, solving its access-level formulation is prohibitively expensive for our full traces, which can contain tens or hundreds of billions of accesses.
For $n$ accesses, \chopt's two-tier graph contains $2n+2$ nodes and up to $4n+2$ edges.
Constructing and storing the graph grows with the number of accesses rather than the number of distinct pages~\cite{CHOPT}.
Prior \chopt evaluations use spatial sampling to reduce the traces solved to around 10 million accesses~\cite{CHOPT}.
Scaling this up to a billion sampled accesses, or a 10 second trace, increases the solution time to around 60 hours, as shown in Figure~\ref{fig:speedup}.
Second, \chopt models migration costs as a constant without considering ongoing migrations or concurrent memory traffic.
That simplification does not hold in practical tiered-memory systems~\cite{ARMS,COLLOID, vulcan_tiering}.
\chopt also restricts migrations to occur immediately before an access.
In real systems, migrations can block concurrent accesses to the page: if a write arrives during migration, it must wait until the copy completes~\cite{HeMem}.
Allowing earlier promotions would let the offline policy move pages before they are needed.
Migrations also interact with demand accesses through bandwidth contention~\cite{MESS}.
A migration during an idle period can be relatively cheap, while the same migration during a memory-intensive phase can stall the application by competing for bandwidth.
The cost of a migration therefore depends on what other accesses and migrations occur nearby in time, so temporally clustered migrations should incur increasing penalties.

Together, these observations point to a gap in characterizing the limits of tiering policies.
This section addresses this gap in three steps: we design an interval-constrained offline-optimal policy under a bandwidth-sensitive migration-cost model, use that policy with a simulator to measure where existing online heuristics fall short, and use those results to motivate a learned online policy.

\subsection{\chomp{} Interval-Constrained Offline-Optimal Policy}
\label{sec:motivation:chomp}

We design \chompfullname, or \chomp, around fixed time steps.
Many modern tiering systems already batch measurements and perform migrations periodically~\cite{HeMem, ARMS, COLLOID, SoarAlto, hybridtier}.
Restricting placement decisions to these intervals lets us compress \chopt's access-level graph and makes room for two extensions: early migrations and bandwidth-sensitive migration costs.

\noindent\textbf{Time-Step Aggregation.}
The blue boxes in Figure~\ref{fig:CHOPT} illustrate how we group the access trace into time steps.
Within each step, we combine repeated accesses to a page into one placement decision, weighted by the total savings from serving those accesses in fast memory.
Because migrations occur only between steps, the two accesses to A in step 2 share one placement decision, as do the two accesses to C in step 3.
The dashed edges mark intermediate migration choices that disappear with this aggregation.
The resulting graph represents pages across time steps rather than individual accesses, greatly reducing its size on traces with many repeated accesses.
As in \chopt, each unit of flow represents one page of fast-memory capacity; the selected flow determines which pages occupy that capacity over time.

\noindent\textbf{Bandwidth-Sensitive Migration Costs.}
The smaller graph gives us the computational budget to encode more accurate costs and migration decisions.
We encode migration interference during graph construction.
For each step, we estimate the bandwidth consumed by application accesses and use the bandwidth/latency profiles supplied by \mess~\cite{MESS} to determine the cost of adding migration traffic.
In addition to fixed migration overheads, such as page-table updates, we encode a shared piecewise-linear interference cost for migrations at that step's boundary.
Its segments have increasing marginal costs: as more pages migrate in the same step, each additional migration becomes more expensive, especially when application traffic already approaches the bandwidth limit.
With these costs encoded before solving, the optimizer balances a page's future access savings against its migration timing and the number of migrations in that step.

\noindent\textbf{Early Migrations.}
We also allow migrations to occur in steps where the page is not directly accessed.
This allows \chomp to migrate pages during an idle step to minimize migration costs during busier steps, while still receiving the full access benefit when the page is eventually accessed.
In the example, C is first accessed in step 3, but \chomp can choose to promote it in step 2 if bandwidth and fast-memory capacity are available.
We also allow the model to select the initial and final placements of pages in fast memory, including an ideal initial placement.

Given the trace, fast-memory capacity, and these costs, \chomp jointly selects placements and migrations across the full trace to minimize migration costs minus fast-memory access savings.
The resulting schedule is optimal within the aggregated-access and migration-cost model, with migrations restricted to interval boundaries; aggregation discards intra-interval access ordering.
We replay this schedule in the simulator to compare it with practical tiering policies.

\noindent\textbf{Model Optimizations.}
We further reduce the graph without changing its optimal solution by pruning migration choices whose future access savings cannot amortize their costs, removing disconnected nodes, and contracting nodes with only one incoming and one outgoing edge.
We then express the graph as a linear program, replacing all flow structures with equivalent constraints.
While the formulation is a linear program, it retains an integral optimum through total unimodularity~\cite{totally_unimodular}.
Applying these same optimizations to \chopt gives \choptpp, which supports approximately 21$\times$ longer traces than unoptimized \chopt but retains access-level decisions.
\chomp's additional scalability comes from time-step aggregation; the smaller graph does not imply solver time is linear in the number of steps.

\input{figures/speedup.tex}

We solve the linear program using the \gurobi solver~\cite{gurobi}.
Our optimizations substantially reduce solver time, and Figure~\ref{fig:speedup} shows the resulting speedups.
Relative to an unoptimized \chopt formulation, \chomp can solve traces that are up to 630$\times$ longer, or 32$\times$ longer than our \choptpp implementation.
This lets \chomp solve traces longer than 10 minutes, sufficient for our evaluation workloads, in about 30 minutes.
Unoptimized \chopt takes around 60 hours to solve a sampled trace covering roughly 10 seconds.
Direct comparison on our full traces is impractical: \chopt's access-level graph is too large to solve, while \chomp's aggregated graph remains tractable.

\input{20-simulator}

%% file: figures/CHOPT.tex
\begin{figure}[t]
\centering
\includegraphics[width=\linewidth,page=9]{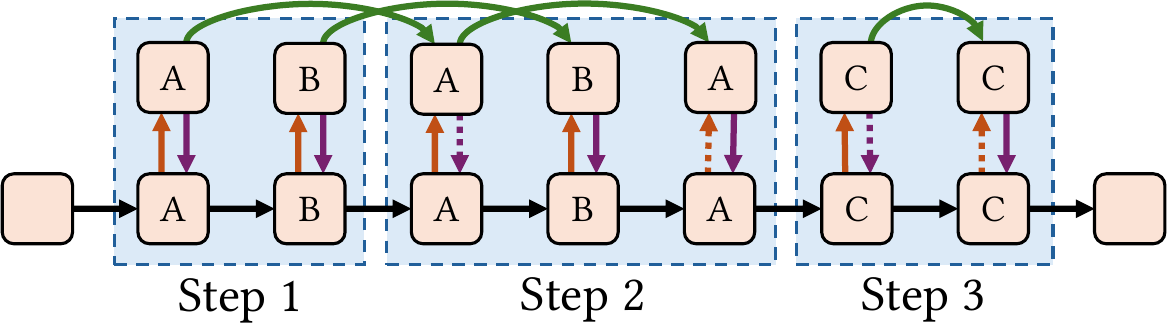}
\Description{The access-level graph for CHOPT with ChOMP's time-step grouping overlaid. Step 1 contains accesses to A and B, step 2 to A, B, and A, and step 3 to C and C. Dashed migration edges become unnecessary when repeated accesses to the same page within a step are aggregated.}
\caption{\chopt's access graph~\cite{CHOPT} with \chomp's time steps (blue boxes). Green edges represent retained pages, orange upward and purple downward edges represent promotions and demotions, respectively. Aggregating repeated accesses removes intermediate migration choices (dashed edges).}
\label{fig:CHOPT}
\end{figure}

%% file: figures/speedup.tex
\begin{figure}[t]
\centering
\includegraphics[width=.8\linewidth]{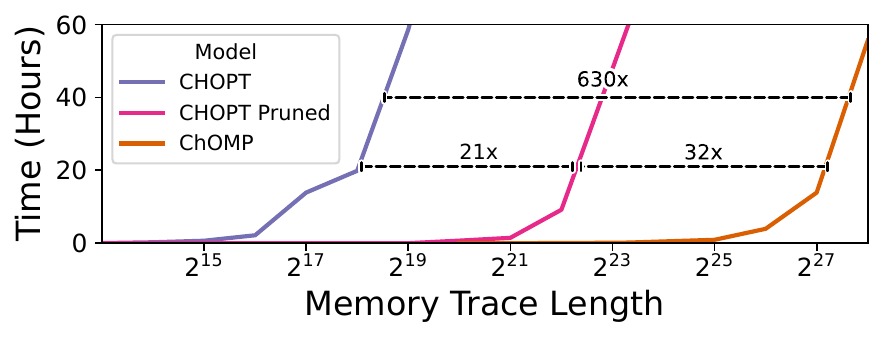}
\caption{Solver time versus trace length for \chopt, \choptpp, and our policy, \chomp.}
\Description{Runtime comparison between \chomp, \chopt, and optimized \chopt on traces of different lengths, showing the improved scalability of \chomp.}
\label{fig:speedup}
\end{figure}

%% file: 20-simulator.tex
\subsection{Simulator}
\label{sec:motivation:simulator}
\input{figures/simulator.tex}

\chomp optimizes a placement-and-migration cost model; it does not directly predict application runtime.
Our trace-driven simulator (Figure~\ref{fig:simulator}) replays arbitrary policies using a common latency, queueing, and bandwidth model without simulating a full CPU.
Because replay captures timing effects beyond \chomp's aggregated cost model, optimality under \chomp's objective does not guarantee minimum simulated runtime, even among interval-based policies.

Inputs are a memory trace, tier sizes, latency--bandwidth profiles, an initialization policy, and a migration policy.
Before replay, we measure the L3 write hit rate and estimate memory parallelism from trace execution rates under all-fast and all-slow memory configurations.

We collect the memory trace~(\figcallout{1}) with Intel PEBS, sampling one in 10,007 events.
Reads use \pebsloadmiss; writes combine \pebsstores with \lthreehitsnoop to estimate write-back traffic.
Because the trace does not identify which stores hit in cache, we approximate caching by capping each page's write count per time step according to the measured L3-miss rate~(\figcallout{2}).
The histogram's red line marks this cap, and its orange portions show the writes removed.
These samples estimate per-page traffic while limiting tracing overhead.

The policy interface~(\figcallout{3}) separates initial placement from migration decisions during replay.
The initialization policy selects the initial contents of fast memory; the tiering policy then chooses promotions and demotions as execution proceeds.
This extensible interface includes \MANTA, \ARMS, \TPP, and a policy that replays \chomp's schedule.
All-fast and all-slow references keep every page in the corresponding tier with migration disabled.

For each demand access or migration transfer, \mess~\cite{MESS}~(\figcallout{4}) returns a latency based on current bandwidth usage.
We use \mess's DDR4 configuration for fast memory, matching our hardware, and measured curves for emulated CXL.
A shared access queue~(\figcallout{5}) models CPU-side timing and backpressure, with the parallelism estimate determining how many operations may be in flight.
Queueing can delay both accesses and migrations; those delays change the traffic seen by subsequent accesses, feeding back into their latency.
Replaying the full trace with this feedback yields the simulated runtime.

\noindent\textbf{Simulator fidelity.}
\label{sec:motivation:simulator-fidelity}
Simulated all-DRAM and all-CXL runtimes differ from hardware measurements by less than 0.1\% on all measured workloads (\xsbench, \prkron, \prtwitter, \bckron, \bctwitter, \multigrid, \faiss, and \duckdb).
These endpoint comparisons show the fit to measured runtimes and support controlled policy comparisons.

\noindent\textbf{Comparing Tiering Policies to \chomp.}
Two behaviors explain heuristic policies' shortfalls.
First, these policies are generally slow to react when a page changes hotness.
A page that transitions from cold to hot may require several control intervals before it crosses the promotion threshold, which delays migration until much of the benefit has already passed.
When a page becomes cold, delayed demotion leaves it occupying fast-memory capacity that could hold currently useful pages.
Figure~\ref{fig:bc-kron-heatmap} illustrates this effect.

Second, many policies encode heuristics for simple patterns that break down on more complex access sequences.
For example, \ARMS requires two consecutive increases in its hotness score to trigger a migration.
This is intended to prevent one-off sequential scans from triggering migrations after the scan has passed, but it can fail on workloads with repeated sequential phases, where the same conservative rule delays useful promotions.

Reactive policies assume that recently measured signals, such as moving averages, are directly informative about future page usefulness.
A learned policy can instead model richer relationships between past access patterns and future page importance over the next migration interval.
We explore these timing differences further in Section~\ref{sec:eval:simulator-evaluation}.
Overall, these observations motivate \MANTA's prediction of future usefulness early enough for migration to pay off.

%% file: figures/simulator.tex
\begin{figure}[t]
\centering
\includegraphics[width=0.9\linewidth,page=1]{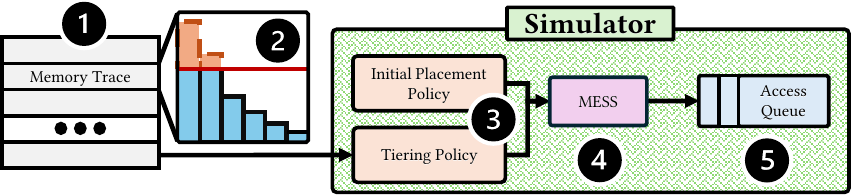}
\caption{Tiering simulator design: \figcallout{1} sampled memory trace, \figcallout{2} per-page write cap approximating caching, \figcallout{3} initial-placement and tiering policies, \figcallout{4} \mess latency model, and \figcallout{5} shared access queue for backpressure.}
\Description{Overview of the tiering simulator with five numbered callouts.
Callout 1 marks the sampled memory trace.
Callout 2 marks a histogram of per-page write counts: a red line caps the counts, and orange portions above it show writes removed to approximate caching.
Callout 3 marks the initial-placement and tiering policies, which determine page placement and migrations.
Callout 4 marks MESS, which models the latency of demand accesses and migration traffic.
Callout 5 marks the shared access queue, which models backpressure and timing effects.}
\label{fig:simulator}
\vspace{-3pt}
\end{figure}

%% file: 40-design.tex
\MANTA replaces heuristic page-hotness estimates with predictions of future usefulness.
Prior systems~\cite{HeMem,ARMS,MEMTIS,hybridtier} improve on Linux-style tiered-memory management with heuristics based on access frequency, recency, or combinations of the two.
These heuristics differ in detail, but they share the assumption that recent behavior is a reliable guide to future behavior.
As shown in Section~\ref{sec:motivation}, that assumption breaks down for workloads whose access patterns evolve continuously or shift abruptly.
\MANTA learns how a page's access history relates to its future accesses, allowing the policy to anticipate changes in page usefulness.

We train a separate model for each application and evaluate one application at a time.
During training-data collection and online deployment, PEBS counters are scoped to the target application, so only its sampled memory accesses contribute to the model's features.
Evaluating one application at a time follows prior learned tiering work (\kleio~\cite{kleio}) and the per-application evaluations of \ARMS, \memtis, \hybridtier, and \nomad~\cite{ARMS,MEMTIS,hybridtier,nomad}.

\subsection{Starting System Design}
\label{sec:design:startingsystem}
We build on \ARMS~\cite{ARMS}, a strong baseline whose hotness-based decisions and migration-control mechanisms let us isolate the benefit of prediction.
We replace its two policy components, hotness estimation and migration-candidate selection, with a learned predictor and a rank-based filter.
Its three mechanism components, access tracking, migration execution, and capacity management, remain unchanged.

\subsection{\MANTA Model}
\label{sec:design:model}
\input{figures/MANTA.tex}
Figure~\ref{fig:MANTA} shows the integration.
The predictor produces a continuous usefulness score for \ARMS's page ranking.
We retain \ARMS's bandwidth-aware throttling, which adjusts migration batch size and concurrency using measured application bandwidth and migration latency~\cite{ARMS}.

\subsubsection{Prediction Objective}
\label{sec:design:objective}
To recognize changing hot sets without overreacting to transient accesses, we predict discounted future page accesses, inspired by discounting in game theory~\cite{game_theory_discounting}:
\begin{equation}
V_p(t) = \sum_{i=1}^{H} \gamma^{i-1} \cdot a_p(t+i)
\label{eq:discounted_value}
\end{equation}
Here, $a_p(t+i)$ counts accesses to page $p$ at step $t+i$, $H$ is the largest future offset available in the trace, and $\gamma\in[0,1]$ controls the relative weight of later accesses.
The sum begins with step $t+1$ and excludes accesses in the current step.
We count all accesses equally, without weighting by type or latency.
Larger values of $\gamma$ emphasize persistent access intensity, while smaller values make the target more responsive to immediate changes.
We use $\gamma=0.99$ to give nearer accesses more weight while still counting accesses far into the trace.
We leave systematic tuning of this value to future work.
This target is the forward-looking counterpart to policies that use moving averages of past accesses~\cite{ARMS}.

\subsubsection{Features}
\label{sec:design:features}
\input{figures/model_features}
Table~\ref{table:model_features} lists the features and their sources.
We aggregate Intel PEBS samples of memory-access events, including LLC misses, into per-page histories.
Moving averages capture temporal patterns~\cite{ARMS}, page age identifies potentially hot new allocations~\cite{TPP}, read/write ratios distinguish access behavior~\cite{HeMem}, and neighboring-page statistics capture spatial correlations~\cite{Coeus,Cronus}.

We inherit \ARMS's 2- and 20-step EWMA horizons, adding 5- and 100-step horizons for intermediate and longer-term history.
The short horizons expose recent changes, while the longer horizons provide context for distinguishing a sustained hot phase from transient activity.
Presenting these histories together lets the model learn their relative usefulness for each application instead of fixing their combination in a heuristic.
This follows the multiple-time-scale approach of \hybridtier~\cite{hybridtier}, with additional precedent for EWMA histories in \memtis and \artmem~\cite{MEMTIS,artmem}.
The \globalavgaccesses feature is a running average of accesses per step over the page's observed history, extending the use of accumulated access statistics in prior systems~\cite{HeMem,MEMTIS,kleio}.

Spatial features use a base neighborhood of eight 2~MB hugepages (16~MB).
\groupabove and \groupbelow each aggregate three consecutive neighborhoods immediately above or below the target neighborhood, covering 48~MB per feature.
Preliminary experiments found three neighborhoods sufficient for spatial context. These sizes are fixed across workloads.

\subsubsection{Data Collection}
\label{sec:design:data_collection}
We collect 10 all-DRAM traces per application with tiering disabled, avoiding migration-induced changes to training accesses.
During collection and deployment, application-scoped PEBS counters exclude operating-system, policy, and other applications' accesses from model features.
We compute each page's target at each step using Equation~\ref{eq:discounted_value}.
For each workload, we fit candidate models on eight traces and evaluate them on two held-out traces during \smac hyperparameter search~\cite{SMAC}.
We then train the deployed model on all 10 traces.

\subsubsection{Virtual Time Steps}
\label{sec:design:virtual_steps}
\ARMS's control loop uses fixed wall-clock intervals, typically 250~ms.
Such windows couple features to execution speed: tiering-induced slowdowns change the number of accesses per window, potentially shifting features outside their training distribution.
The same application phase can therefore produce different feature magnitudes during all-DRAM training and tiered execution, even when its underlying access pattern is unchanged.
We instead advance virtual time after a fixed number of the application's PEBS samples.
This measures access history in units of application progress and reduces the feedback from earlier placement decisions into the feature representation.
To balance temporal resolution against bookkeeping and inference costs, we measure samples per 250~ms during all-DRAM execution and round to the nearest $10^{k/2}$ for integer $k$.
This yields 10,000 samples per step for most workloads, 3,162 for \faiss~\cite{faiss}, and 100,000 for \duckdb.
Section~\ref{sec:eval:policy-ablations} evaluates this choice.

\subsubsection{Model Architecture}
\label{sec:design:model_architecture}
Scoring tens of thousands of pages multiple times per second constrains the models that are practical for online tiering.
We use a \lightgbm random-forest regressor~\cite{LightGBM} for low-latency inference on structured features.
Using \lleaves~\cite{lleaves}, we compile the trained model to standalone native code, avoiding the full \lightgbm runtime in the online path.
Compilation reduces inference time by an order of magnitude to under 20~ms for more than 10,000 pages, enabling predictions every control interval.

\subsubsection{Parameters and Tuning}
\label{sec:design:tuning}
\smac~\cite{SMAC} searches a predefined regressor-hyperparameter space using a small data subset.
It tunes only model hyperparameters, which affect accuracy, size, and inference cost.
Features, EWMA horizons, and neighborhood sizes are fixed across workloads.

\subsubsection{Migration Candidate Filter}
\label{sec:design:migration_filter}
\input{figures/MANTA_filter.tex}
At deployment, we turn the model's predictions into migration decisions in two steps: average each page's predictions over the last 10 control intervals to stabilize its score, then filter the resulting page ranking to select promotions and demotions.
Smoothing reduces noise in the predicted scores, but can delay reactions to abrupt changes in access patterns.
Additionally, a high predicted score identifies a useful page, but does not by itself justify moving it immediately because migration consumes bandwidth and can interfere with application accesses.

\ARMS's cost-benefit mechanism couples migration decisions to the momentum of its hand-engineered hotness scores~\cite{ARMS}.
It estimates migration benefit from reactive scores and hot age, then compares that benefit with observed migration costs.
These inputs do not map directly onto \MANTA's predictions of future usefulness.

We therefore use a simpler rank-based filter that operates directly on the ordered hot and cold sets shown in Figure~\ref{fig:MANTA_filter}.
The hot set contains the highest-ranked pages up to the fast-memory capacity, and the cold set contains the remaining pages below that capacity boundary.
A page in slow memory is a promotion candidate only if it lies in the top half of the hot set.
A page in fast memory is a demotion candidate only if it lies in the bottom half of the cold set.
The hatched region in the figure covers the excluded halves adjacent to the capacity boundary.
These pages stay in place that interval.

We do this because the model predictions can be noisy near the migration threshold, where small score fluctuations can otherwise cause unnecessary page movement.
Together, prediction smoothing and rank filtering reduce churn and keep the system from spending migration bandwidth on pages whose predicted ordering is unstable from one interval to the next.
Using ranks avoids calibrating a migration-cost multiplier for each workload; the rank threshold and prediction-smoothing window remain fixed across the workloads we evaluate.
Section~\ref{sec:eval:cost-ablation} compares this rank threshold with an \ARMS-like cost-benefit check applied to our learned scores, using a fixed migration-cost scaler for each configuration.

\subsubsection{Model Validity and Retraining}
\label{sec:design:retraining}
We evaluate applications individually, without co-running workloads.
Under co-location, each application would need its own model and features while sharing memory and migration bandwidth.
Application-scoped counters separate feature streams, but do not isolate access timing or performance from other tenants' contention.
\ARMS throttling limits migration pressure without providing cross-tenant bandwidth allocation, fairness, or performance isolation; evaluating these scenarios remains future work.

Model reuse depends on whether new access behavior remains represented by the training traces.
A new dataset or input size does not necessarily require retraining; we evaluate transfer across datasets, sizes, and execution lengths in Section~\ref{sec:eval:sensitivity}.
Changes to code, control flow, input type, dataset size or skew, thread count, or algorithm parameters can nevertheless shift access behavior.
After such changes, the operator must validate predictions and runtime features on representative traces.
Retraining is required when features fall outside the training distribution, or prediction quality or application performance degrades.
Validation must therefore assess decision quality as well as feature distributions: a model can lose its usefulness even when no individual feature gives an obvious drift signal.

We use an operator-managed lifecycle: collect new PEBS traces, rebuild the model off-host, and replace it.
Section~\ref{sec:eval:model-building-cost} quantifies rebuilding cost.
Monitoring input distributions and decision quality, as proposed by Guardrails for the OS~\cite{os_guardrails}, could trigger heuristic fallback or retraining.
Automatic drift detection and online updates remain future work.

%% file: figures/MANTA.tex
\begin{figure*}
\centering
\includegraphics[width=.88\linewidth,page=5]{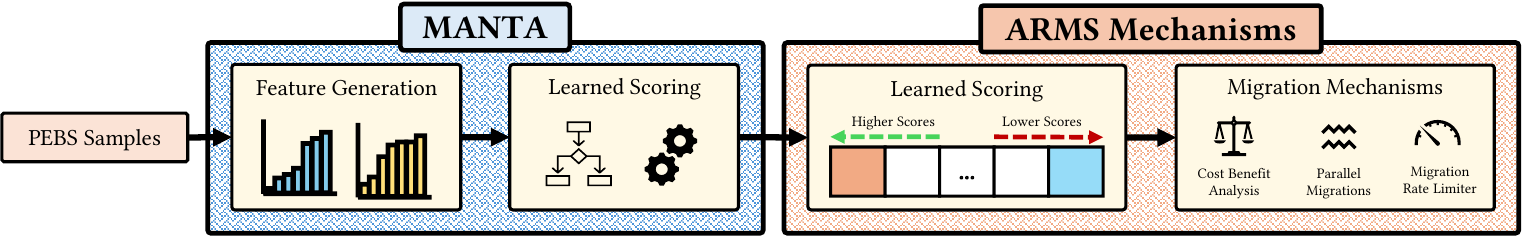}
\caption{\MANTA integration: blue components are added or modified, orange provided by the kernel, and yellow inherited from \ARMS.}
\Description{System Overview of \MANTA.}
\label{fig:MANTA}
\end{figure*}

%% file: figures/model_features.tex
\begin{table*}[t]
\centering
\small
\begin{tabular}{@{}>{\raggedright\arraybackslash\leavevmode}p{0.20\textwidth}%
>{\raggedright\arraybackslash\leavevmode}p{0.46\textwidth}%
>{\raggedright\arraybackslash\leavevmode}p{\dimexpr0.33\textwidth-4\tabcolsep\relax}@{}}
\toprule
\textbf{Feature} & \textbf{Description} & \textbf{Prior source(s)} \\
\midrule
\ewmatwo, \ewmafive, \ewmatwenty, \ewmahundred & Exponentially weighted moving averages of page accesses using several target window sizes & {\footnotesize\ARMS~\cite{ARMS}; \hybridtier~\cite{hybridtier}; \memtis~\cite{MEMTIS}; \artmem~\cite{artmem}} \\
\hline
\globalavgaccesses & Running average of the page's accesses per step over its observed history & {\footnotesize\hemem~\cite{HeMem}; \memtis~\cite{MEMTIS}; \hybridtier~\cite{hybridtier}; \kleio~\cite{kleio}} \\
\hline
\pageage & The age of the page in steps & {\footnotesize\hemem~\cite{HeMem}; \memtis~\cite{MEMTIS}; \TPP~\cite{TPP}; \ARMS~\cite{ARMS}; \hybridtier~\cite{hybridtier}; \nomad~\cite{nomad}} \\
\hline
\rratiotwenty, \rratiohundred & Long-term read/write ratios for the page & {\footnotesize\hemem~\cite{HeMem}; \vulcan~\cite{vulcan_tiering}} \\
\hline
\groupabove, \groupbelow & Average access count over three consecutive eight-page neighborhoods immediately above or below the target neighborhood (48~MB per feature) & {\footnotesize\cronus~\cite{Cronus}; \coeus~\cite{Coeus}} \\
\hline
\nearbypages & Average access count in the eight-page (16~MB) neighborhood containing the hugepage & {\footnotesize\cronus~\cite{Cronus}; \coeus~\cite{Coeus}} \\
\hline
\nearbyewmavariance & Variance of the access-count EWMA within the eight-page (16~MB) neighborhood containing the hugepage & {\footnotesize\cronus~\cite{Cronus}; \coeus~\cite{Coeus}; \ARMS~\cite{ARMS}} \\
\bottomrule
\end{tabular}
\caption{Runtime features used by the current predictive model.}
\label{table:model_features}
\end{table*}

%% file: figures/MANTA_filter.tex
\begin{figure}
\centering
\includegraphics[width=.7\linewidth,page=6]{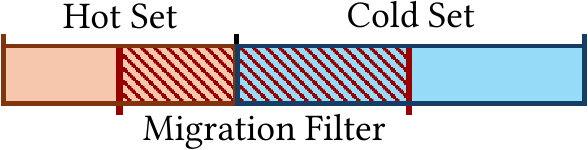}
\caption{\MANTA's rank filter replaces \ARMS's hotness-dependent filters.}
\Description{Migration filter added by \MANTA, which replaces \ARMS's boundary filters that depend on \ARMS's original hotness estimator.}
\label{fig:MANTA_filter}
\end{figure}

%% file: 50-evaluation.tex
\input{figures/6_2_c220g5_eval.tex}

In this section, we compare \MANTA against existing tiering policies.
We ask four questions: (1) How does \MANTA perform across workloads, kernels, and memory technologies? (2) How well does a model transfer to new inputs and longer executions? (3) How much of the gap to \chomp does it recover? (4) What are its prediction accuracy, overhead, and sensitivity to individual design choices?
We use two hardware configurations.
The first is a \cloudlabnode CloudLab~\cite{Cloudlab} machine that emulates CXL using a downclocked remote NUMA node, similar to prior work~\cite{ARMS, COLLOID, SoarAlto}.
The second is a self-hosted Optane-based system.

\noindent\textbf{Workloads.}
We evaluate workloads sensitive to memory tiering, running one application at a time without competing background workloads or multi-tenancy.
For graph analytics workloads, we use Betweenness Centrality (\bc)~\cite{GapBS} and PageRank (\pr)~\cite{PageRank} on the Kron and Twitter graphs (\bckron, \bctwitter, \prkron, and \prtwitter).
For scientific computing, we use \xsbench~\cite{XSBench}, an application for Monte Carlo neutron transport, and \multigrid (Class D)~\cite{nasa_npb}, an application that benchmarks multigrid solvers.
For vector databases, we use \faiss~\cite{faiss} on OpenAI text embeddings~\cite{openai_embeddings}.
Finally, for databases, we use \duckdb~\cite{duckdb} running the TPC-H SF100~\cite{tpch} workload.
We use all eight workloads on emulated CXL; Optane experiments omit \duckdb and \multigrid because both perform extremely poorly when pages are placed in Optane (Section~\ref{sec:eval:optane}).
The policy ablations use \bctwitter and \prtwitter to contrast workloads with larger and smaller gains from prediction (Section~\ref{sec:eval:policy-ablations}).
The main workloads have memory footprints of 13--27~GB; the larger-input experiment is described separately in Section~\ref{sec:eval:dataset-scale-transfer}.
The 4--10~GB fast tier is smaller than each workload's full footprint; increasing its capacity allows more frequently accessed pages to remain in fast memory.
\MANTA uses 2~MB pages, and the capacities on the x-axes denote fast-tier memory, not total workload size.
We vary fast-memory capacity because it affects policy performance~\cite{MEMTIS, hybridtier}.
Each tiering-policy workload, memory capacity, and kernel combination is repeated three times.
Per-workload error bars show the minimum and maximum per-run speedups.

\noindent\textbf{System Setup.}
Across evaluations, we use the same workloads and, where possible, comparable kernel versions so that differences reflect tiering policy rather than unrelated system changes.
For the emulated-CXL experiments, we run on a CloudLab \cloudlabnode node~\cite{Cloudlab} with two NUMA regions, each containing a Xeon Silver 4114 CPU and 192 GB of DRAM.
One NUMA region serves as fast memory and the other as slow memory.
To make the remote NUMA region more representative of CXL, we downclock its uncore frequency to 700 MHz, increasing access latency using the same methodology as prior work~\cite{ARMS}.
The baseline implementations use different patched kernels, as detailed in Section~\ref{sec:eval:sim-cxl}.
Our Linux 6.2 results use \hybridtier's patched \texttt{6.2.0-hybridtier+} kernel for \MANTA, \ARMS, and \hybridtier.

We then repeat the \ARMS-versus-\MANTA comparison on Linux \texttt{6.18.1} to evaluate performance on a newer kernel.
Both kernel versions use the same Spectre mitigations.
The all-DRAM and all-CXL references include both kernel series.

Finally, we repeat the experiments again on our Optane system with the Linux \texttt{6.18.1} kernel.
Our Optane system uses an \xeongold CPU, with six channels of \hynixdram as fast memory, and \optanepmem devices as slow memory in the same NUMA region, providing 135~GB of remote memory.

\noindent\textbf{Model-Building Cost.}
\label{sec:eval:model-building-cost}
Figure~\ref{fig:model-characterization} reports measured model build times of 455--883~s across the eight workloads, rounded to whole seconds.
We sum hyperparameter search, full-dataset construction, finalist evaluation, final test evaluation and model saving, and model compilation from each model's timing log.
This sum excludes trace collection, input discovery, dependency loading, and streaming/reservoir sampling.
This offline cost is paid at model creation and retraining (Section~\ref{sec:design:retraining}), not on each application run.

\subsection{Comparison with Baselines on Emulated CXL}
\label{sec:eval:sim-cxl}

We first compare \MANTA against all selected baselines using emulated CXL.
We use \ARMS, \TPP~\cite{TPP}, \hybridtier~\cite{hybridtier}, \nomad~\cite{nomad}, and \memtis~\cite{MEMTIS} as our baselines.
We omit a separate \TPP{}+\colloid experiment and use \TPP as the baseline for this single-application evaluation.
This is because \TPP is roughly equivalent to \TPP{}+\colloid in the single-application regime where the default tier remains lower-latency under load: the \colloid paper's default GUPS experiment reports matching performance at $0\times$ injected memory-interconnect contention~\cite{COLLOID}.
\TPP, \hybridtier, \nomad, and \memtis are kernel implementations.
\TPP uses Linux \texttt{5.15.0+} with TPP patches, and \hybridtier uses Linux \texttt{6.2.0-hybridtier+}.
\nomad uses Linux \texttt{5.13.0-rc6+} with NOMAD patches and the \texttt{async\_promote} module.
\memtis uses Linux 5.15.19 with HTMM patches.
\ARMS and \MANTA share \hybridtier's kernel.

Figure~\ref{fig:6_2_c220g5-eval} shows our results.
\MANTA has the highest geometric-mean speedup of all evaluated systems at every capacity.
The strongest aggregate baseline is \ARMS at 4 and 6~GB and \memtis at 8 and 10~GB; \MANTA outperforms these baselines by approximately 1.12$\times$, 1.07$\times$, 1.03$\times$, and 1.03$\times$, respectively.
Against \memtis alone, the corresponding speedups are approximately 1.28$\times$, 1.11$\times$, 1.03$\times$, and 1.03$\times$.
Against \ARMS, the geometric-mean speedups at 4, 6, 8, and 10~GB are approximately 1.12$\times$, 1.07$\times$, 1.08$\times$, and 1.08$\times$, respectively.
Individual workloads can still favor another system, as the panels show.
For \bc, \MANTA achieves 1.03--1.24$\times$ speedup over \ARMS on \bckron and 1.05--1.16$\times$ on \bctwitter.
It outperforms every baseline on \bckron, including 1.08--1.59$\times$ gains over \memtis.
On \bctwitter, \MANTA outperforms \memtis by 1.28$\times$ and 1.22$\times$ at 4 and 6~GB, respectively, while \memtis is 1.03$\times$ and 1.05$\times$ faster at 8 and 10~GB.

On \xsbench, \MANTA achieves 1.22--1.25$\times$ speedup over \ARMS across capacities.
It also outperforms \memtis by 1.28$\times$ at 4~GB, but \memtis leads by 1.07--1.08$\times$ at 6--10~GB.
\MANTA and \TPP are within 1\% at 8~GB; \MANTA leads \TPP at the other capacities.
On \faiss, \MANTA outperforms every plotted baseline, with gains of 1.02--1.17$\times$ over \memtis.
\duckdb has a different ordering: \memtis is 1.11$\times$ and 1.05$\times$ faster than \MANTA at 4 and 6~GB, respectively; the two are within 1\% at 8~GB, and \MANTA leads by 1.04$\times$ at 10~GB.
\MANTA nevertheless outperforms \ARMS on \duckdb at every capacity, by 1.06--1.13$\times$.

For \pr, \MANTA achieves 1.04--1.10$\times$ speedup over \ARMS on \prkron and 1.01--1.06$\times$ on \prtwitter.
It outperforms \memtis by 1.02--1.75$\times$ on \prkron, although \TPP or \nomad can lead at larger capacities.
On \prtwitter, \MANTA leads \memtis by 1.61$\times$ at 4~GB, while \memtis is 1.04--1.08$\times$ faster at 6--10~GB.
For \multigrid, \MANTA outperforms \memtis by 1.10--1.31$\times$ and remains within 1\% of \ARMS at 6~GB, but \hybridtier is 1.04--1.10$\times$ faster than \MANTA at 6--10~GB.

Kernel behavior accounts for much of this performance difference between \MANTA and other baseline systems.
There is one exception to this in DuckDB, which benefits from initial placement policies that \ARMS and \MANTA do not have.
Further differences are explored in Section~\ref{sec:eval:simulator-evaluation}.
\MANTA and \ARMS share \hybridtier's \texttt{kswapd} changes in 6.2, while \memtis, \TPP, and \nomad use their own patched kernels.
Additionally, these systems implement their own memory-management mechanisms inside the kernel, which enables initial placement policies which we cannot support, as well as base page migrations.
Comparing \MANTA with \ARMS on the same kernel holds these kernel mechanisms constant while assessing \MANTA's policy changes.
Figure~\ref{fig:sim_eval} complements the measured system comparisons with a simulated \chomp reference.
We next repeat the \MANTA-versus-\ARMS comparison on Linux 6.18 to test whether \MANTA's gains persist on a newer kernel.

\subsection{Comparison on a New Kernel}
\label{sec:eval:new_kernel}

\input{figures/6_18_c220g5_eval.tex}

We next rerun \MANTA and \ARMS on Linux 6.18 to examine how the newer kernel affects their performance and whether \MANTA's gains persist, as shown in Figure~\ref{fig:6_18_c220g5-eval}.
We reuse the Linux 6.2 models unchanged, without retraining or retuning.
We attribute the main performance difference between these kernels to additional \texttt{kswapd} activity from \hybridtier's 6.2 patches in our setup, as well as improvements in memory management and CPU scheduling in Linux 6.18.
Across the eight workloads, \MANTA achieves approximately 1.08$\times$, 1.03$\times$, 1.06$\times$, and 1.07$\times$ geometric-mean speedup over \ARMS on Linux 6.18 at 4, 6, 8, and 10~GB, respectively.
Because the other baseline implementations do not support this kernel, the direct comparison is between \MANTA and \ARMS; we also include their Linux 6.2 performance values and the best-performing baseline not already included in the figure for each workload and capacity.
The ``Best Baseline'' bars select the lowest mean runtime among \hybridtier, \nomad, \TPP, and \memtis separately for each workload/capacity combination, excluding \ARMS and \MANTA.

\MANTA retains its aggregate advantage on the newer kernel, with sustained gains across capacities on several workloads.
On \duckdb, it delivers approximately 1.11$\times$ speedup over \ARMS at every capacity.
On \prtwitter, its 1.06--1.17$\times$ gains exceed the 1.01--1.06$\times$ range on Linux 6.2.
\xsbench retains a 1.07--1.16$\times$ advantage, although smaller than its 1.22--1.25$\times$ gains on Linux 6.2.

The capacity trends differ by workload.
For \faiss, \MANTA's speedup grows to 1.08$\times$ and 1.21$\times$ at 8 and 10~GB, respectively; at 4--6~GB, the systems are within 1.3\%.
Similarly, \multigrid improves from a near-tie at 4~GB to 1.03$\times$, 1.09$\times$, and 1.12$\times$ at 6, 8, and 10~GB, with no regression against \ARMS.
For \bc and \prkron, the gains are concentrated at the smallest fast-tier capacity: at 4~GB, \MANTA achieves 1.09$\times$ on \bckron, 1.18$\times$ on \bctwitter, and 1.06$\times$ on \prkron.
At 6--10~GB, these three workloads remain close to \ARMS, which leads by at most 3.1\%.

\subsection{Comparison on Optane}
\label{sec:eval:optane}
\input{figures/optane_eval.tex}

We use Optane to represent slow memory with different latency and bandwidth characteristics from emulated CXL.
The Optane experiment uses a smaller workload subset: the four graph workloads, \xsbench, and \faiss.
We remove \duckdb and \multigrid as they perform poorly when any amount of memory is placed in Optane.
We compare only against \ARMS because the Optane machine does not support the kernel changes required by the other baselines.
Figure~\ref{fig:optane_eval} reports speedup relative to all-Optane execution, with all-DRAM execution as a reference; higher values indicate better performance.
Across the six workloads, \MANTA achieves approximately 1.69$\times$, 1.13$\times$, 1.14$\times$, and 1.09$\times$ geometric-mean speedup over \ARMS at 4, 6, 8, and 10~GB, respectively.
The largest plotted speedup ratio is 5.63$\times$ on \prtwitter at 4~GB; \prkron reaches 2.00$\times$ at the same capacity.

The clearest gains occur for \bc, \xsbench, and \pr under memory pressure. As fast-memory capacity grows, \MANTA's advantage on \pr narrows, with regressions of up to about 4\% relative to \ARMS at some larger capacities. Across the four capacities, \faiss's mean runtimes differ by only 0.2--1.7\%, so these measurements show little benefit from changing the policy on that workload.

\subsection{Generalization to New Settings}
\label{sec:eval:sensitivity}

\subsubsection{Longer Workload Runs}
\label{sec:eval:long_runs}

\input{figures/long_eval.tex}

Figure~\ref{fig:long_eval} compares \MANTA and \ARMS on longer executions of \bctwitter and \prtwitter on Linux 6.18 across four memory capacities.

We extend both workloads to 100 iterations, 10$\times$ the training-run iteration count, while retaining the fixed models used in Figure~\ref{fig:6_2_c220g5-eval}.
We measure new all-DRAM and all-CXL references for these longer executions.

For the comparisons below, we divide \ARMS's mean runtime by \MANTA's mean runtime over three successful runs per configuration; ratios above one favor \MANTA.
At 4, 6, 8, and 10~GB, respectively, these ratios are approximately 1.086$\times$, 1.069$\times$, 1.004$\times$, and 0.999$\times$ for \bctwitter.
Runtime is 7.9\%, 6.5\%, and 0.4\% lower at 4, 6, and 8~GB, respectively, and 0.1\% higher at 10~GB.
Thus, \bctwitter retains an advantage at the smaller capacities, although its 4~GB advantage narrows from approximately 1.18$\times$ in the shorter Linux 6.18 run.
For \prtwitter, the longer-run ratios are approximately 0.984$\times$, 0.987$\times$, 0.972$\times$, and 0.997$\times$: runtime increases by 1.7\%, 1.3\%, 2.9\%, and 0.3\%, respectively.
Its approximately 1.06--1.17$\times$ advantage in the shorter Linux 6.18 runs therefore does not persist.

The workloads differ in how their access patterns evolve: \bc remains relatively stable throughout the trace, whereas \pr's patterns drift as execution progresses~\cite{Presto}.
We interpret \bctwitter's gains at smaller capacities and near-ties at larger capacities as consistent with the model being able to generalize from the training run to the longer execution.
For \prtwitter, a shift in the access distribution during later phases can reduce the representativeness of the training data and explain the loss of the original run's advantage.

These results illustrate the model-validity limits in Section~\ref{sec:design:retraining}: extending execution can preserve useful predictions, but later phases outside the training distribution require validation and potentially retraining.
The model remains fixed throughout this experiment; it does not adapt through online retraining.
Section~\ref{sec:eval:model-characterization} reports per-step prediction-quality percentiles; these do not measure the distribution of interval-level runtime slowdowns.

\subsubsection{Cross-Dataset Transfer}
\label{sec:eval:cross-dataset-transfer}

For both \bc and \pr, we deploy the model trained on Twitter on the Kron graph, and vice versa.
\input{figures/cross_dataset_eval.tex}
Figure~\ref{fig:cross-dataset-eval} compares these transferred models with \ARMS and the dataset-specific models from Figure~\ref{fig:6_2_c220g5-eval} on Linux 6.2, using speedup over all-CXL execution.
The transferred models generally match or outperform \ARMS while retaining the advantage of the dataset-specific models.
The occasional regressions, including \prtwitter at the largest capacity, show that the benefit is not uniform across configurations.
These results support reuse on new datasets of the trained applications, subject to validation when access behavior changes (Section~\ref{sec:design:retraining}).

\subsubsection{Dataset-Scale Transfer}
\label{sec:eval:dataset-scale-transfer}

\input{figures/dataset_scale_eval.tex}

We scale the Kron graph from 20 to 80~GB while retaining the same degree parameter, and reuse the earlier \bckron and \prkron models for their respective workloads.
Figure~\ref{fig:dataset-scale-eval} compares these models with \ARMS on Linux 6.18.
The models, features, virtual-step size, and 2~MB page size remain unchanged, with no retraining or retuning.
The 80~GB input size denotes the graph data.
The measured resident footprints are close to 80~GB, with only marginal additional overhead.

For \bc, transfer achieves speedups over \ARMS of 1.11$\times$ and 1.09$\times$ at 4~GB and 6~GB, respectively, with marginal speedups at higher capacities.
For \pr, the largest benefit occurs at 4~GB with a 1.21$\times$ speedup, where there is the most contention for fast-tier memory.
The policies converge to be roughly equivalent at larger fast-tier capacities.

Transfer broadly preserves the benefits seen on the smaller graph, particularly under memory pressure.
The larger working set makes \pr more difficult for \ARMS at the smallest capacity; with more fast memory, the policies converge.

These results support reuse within the same graph family, even as the graphs grow in capacity.
Input growth alone need not trigger retraining; degraded predictions or performance should prompt validation and, where necessary, retraining (Section~\ref{sec:design:retraining}).

\subsection{Simulator Evaluation}
\label{sec:eval:simulator-evaluation}

\input{figures/sim_eval.tex}

We use the simulator to measure how much of the runtime gap between \ARMS and the interval-constrained \chomp reference is closed by \MANTA.
Figure~\ref{fig:sim_eval} compares measured runtimes for \MANTA, \ARMS, and \hybridtier with simulated \chomp and all-DRAM references across eight workloads and 2-8~GB of DRAM.
All bars are normalized to the simulated all-CXL runtime for the corresponding workload.
Section~\ref{sec:motivation:simulator} describes the calibration and reference construction.

\MANTA recovers much of the available headroom by preventing unnecessary page churn and making useful placements sooner.
Its predictions allow it to promote pages without waiting for their reactive hotness scores to ``warm up,'' while retaining useful pages avoids repeated eviction and promotion.
On \bctwitter at 2~GB, \MANTA saves 18.0~s of the 21.8~s gap to \chomp, recovering 82\% and delivering a 1.18$\times$ speedup over \ARMS.
On \prtwitter at 6 and 8~GB, it saves 47.7 and 27.7~s out of gaps of 60.1 and 35.4~s, recovering 79\% and 78\%, respectively.
At 8~GB, \MANTA finishes only 7.8~s, or 2.9\%, slower than the \chomp reference.
These results show that predictive placement can remove most of the loss incurred by waiting for reactive scores to catch up.

Capacity determines how much additional improvement is possible.
For \bckron, \prkron, and \bctwitter at 4--8~GB, \ARMS already captures most of the benefit of fast memory, leaving comparatively little headroom for \MANTA to recover.
The two policies' runtimes differ by at most 2\% on the Kron workloads and 3.4\% on \bctwitter, including small regressions for \MANTA.
For example, at 8~GB on \bctwitter, \ARMS and \MANTA finish in 97.3 and 99.0~s, compared with 91.7~s for \chomp.
Here, replacing the ranking policy has limited scope to improve performance.
On \multigrid, however, \MANTA's advantage grows with capacity: it saves just 0.3~s at 2~GB, but 44.7~s of a 77.2~s gap at 8~GB, recovering 58\%.
This pattern is consistent with better page selection becoming more useful as enough fast memory becomes available to retain the selected pages.

\faiss illustrates both the benefit of reducing churn and the remaining cost of migration delay.
Churn makes \ARMS's placements ineffective, while \MANTA's predictions improve which pages remain in fast memory.
At 8~GB, \MANTA saves 96.0~s of the 173.3~s gap, recovering 55\% and achieving a 1.22$\times$ speedup over \ARMS.
Nevertheless, \MANTA still trails \chomp by 77.3~s because identifying the right pages does not place them in DRAM immediately.
At 2--4~GB, the two policies remain within 1.3\% of each other despite remaining headroom, so the benefit is concentrated at larger capacities.

\duckdb exposes a different limitation: frequent deallocations and reallocations repeatedly discard useful placements.
\MANTA allocates new pages in the far tier and must subsequently identify and promote them, so repeated allocation forces it to pay this delay again.
It still outperforms \ARMS by about 1.11$\times$ across capacities, recovering 39--55\% of the gap, but much of the allocation-related loss remains.
At 2~GB, \MANTA saves 48.0~s over \ARMS yet remains 39.7~s above \chomp.
Its 443.3~s runtime is just faster than the 445.9~s all-CXL reference; \ARMS and \hybridtier remain slower than all-CXL execution.
Thus, this case calls for better initial placement as well as better hotness prediction.

\xsbench presents the same placement challenge primarily once, during initial allocation: performance depends on how quickly the useful pages reach the right tier.
\MANTA avoids waiting for hotness scores to build up and saves 26.3~s of the 44.3~s \ARMS-to-\chomp gap at 2~GB, recovering 59\%.
Across capacities, its remaining gap stays near 16--18~s, consistent with a placement delay that extra DRAM alone does not eliminate.
\MANTA also outperforms \hybridtier at every capacity on \xsbench, \duckdb, and all four graph workloads, although \hybridtier retains a small advantage on \multigrid at 4--8~GB.

\subsection{Model Characterization}
\label{sec:eval:model-characterization}

\input{figures/model_characterization.tex}

Figure~\ref{fig:model-characterization} summarizes prediction quality, per-page and per-step processing time, and model build time across eight workloads.
After building each model, we collect separate online traces specifically for this evaluation; they are never used for training or tuning.

Model building takes roughly 8--15 minutes for the stages in Section~\ref{sec:eval:model-building-cost}.
Hyperparameter optimization (HPO) takes about six minutes, with final model building and evaluation accounting for most of the remaining time.
Compilation adds less than 1\% of the overall time.

Feature generation and inference measure the per-page costs of constructing model inputs and computing predictions, respectively.
Total reports processing time per timestep, including feature generation and inference; the timing columns report means and, for inference and total, P95 values.
$R^2$ measures prediction fit to discounted future reward with discount 0.99, pooling all finite page observations (Section~\ref{sec:design:objective}).
We evaluate the top 1,000 pages (roughly 2~GB) because differences in hot-page selection are most apparent under tight capacity.
The system-level comparisons in Sections~\ref{sec:eval:sim-cxl} and~\ref{sec:eval:new_kernel} show that \MANTA's runtime gains generally persist at larger fast-tier capacities.
Capture is the reward of the selected pages as a percentage of the oracle top-1,000 reward.
Collected is the percentage of each timestep's hits that fall on the selected pages.
Recall is the percentage of the oracle's hottest positive-reward pages included in the selection, using all such pages when fewer than 1,000 exist.
Capture measures how much reward the selection collects, while recall measures which pages it selects; higher capture can therefore coexist with lower recall.
Both metrics average equally over steps with positive reward.

Turnover is the percentage of selected pages that changed between consecutive timesteps.
It measures churn in the hot-page set and thus potential migration work: lower turnover is desirable because migrations are expensive.
Actual migrations also depend on the deployed rank filter (Section~\ref{sec:design:migration_filter}).
P5 is the fifth percentile of per-step capture or recall, highlighting weaker prediction intervals; P95 is the ninety-fifth percentile of the corresponding timing or turnover measurement.
These characterize processing overhead and selection stability, not interval-level application slowdowns or memory-access tail latency.

\MANTA's mean total processing time ranges from 16.026 to 45.697~ms per step, with P95 values of 16.975--61.352~ms; \ARMS's P95 total is 5.472--22.911~ms.
\MANTA's P95 inference cost is 0.645--2.359~$\mu$s per page.

Across these traces, \MANTA's $R^2$ ranges from 0.553 to 0.966.
It achieves higher Capture@1k and lower Turnover@1k than \ARMS on all eight workloads, and higher Recall@1k on seven of eight.
Only \faiss has lower recall, despite higher capture.
\MANTA thus generally improves hot-page selection while achieving substantially lower churn across every workload, reducing the potential need for costly migrations.

The percentile results show that these benefits extend beyond the mean.
\MANTA improves P5 capture on all eight workloads and P5 recall on seven, with \faiss again the exception.
For example, P5 capture rises from 70.95\% to 99.74\% on \xsbench and from 82.75\% to 97.37\% on \bckron.
P95 turnover is lower on seven workloads, falling from 24.87\% to 1.50\% on \bckron; \xsbench is tied at 19.30\% despite lower mean turnover.
Thus, \MANTA generally maintains better selection quality in weaker intervals while also reducing high-churn episodes.

\subsection{Ablation Studies}
\label{sec:eval:ablations}

\subsubsection{Cost Ablation}
\label{sec:eval:cost-ablation}

\input{figures/6_18_c220g5_cost_ablation_twitter.tex}

Figure~\ref{fig:6_18_cost_ablation_c220g5-eval} compares \MANTA's rank-based migration filter (Section~\ref{sec:design:migration_filter}) with an \ARMS-like cost-benefit check on \bctwitter and \prtwitter.
The variants retain \MANTA's learned predictor and replace its rank filter with a check that requires predicted benefit to exceed a scaled migration cost.
Lower scalers admit more candidates, while higher scalers can reject useful promotions.

On \bctwitter at 4~GB, scalers of 0.125--0.25 increase runtime by 10.03--15.92\% relative to the default rank filter; higher scalers recover most of this loss.
On \prtwitter, however, scalers of 0.5--1.5 increase runtime by 7.14--9.71\% at 8--10~GB.
Thus, a setting that works well for one workload and capacity can hurt another.
We tested all evaluated workloads and capacities on each platform. The others show similar tradeoffs or weaker sensitivity, so we present these two workloads to illustrate the variation.
The rank filter reduces the need to tune a migration-cost scale for each workload and capacity.

\subsubsection{Policy Ablations}
\label{sec:eval:policy-ablations}

\input{figures/optane_ablation_eval.tex}

Figure~\ref{fig:optane_non_model_ablation} tests virtual steps and the rank filter on \bctwitter and \prtwitter with 4~GB and 6~GB of fast memory on Optane, contrasting gains under tight capacity with smaller differences from \ARMS at higher capacity.
The panels report speedup over all-Optane execution; higher is better.
Runtime ratios use arithmetic mean runtimes for each workload and capacity; the bars average per-run speedups.

First, we remove the virtual-step data collection and retrain a model that uses a fixed 250~ms step size.
Without virtual steps, mean runtime increases by 15.7\% and 30.2\% on \bctwitter at 4 and 6~GB, respectively, and by 33.8\% and 5.8\% on \prtwitter.
At 4~GB, the fixed-step variant still achieves 1.28$\times$ and 4.41$\times$ speedups in mean runtime over \ARMS on \bctwitter and \prtwitter, respectively; at 6~GB, it takes 24.9\% and 8.4\% longer than \ARMS.
Thus, virtual steps improve performance in all four configurations, but removing them does not eliminate every benefit of the learned policy.

Second, we apply the rank filter, designed for \MANTA's predicted scores, to \ARMS's reactive scores to test whether filtering alone explains \MANTA's gains.
On \bctwitter, filtered \ARMS takes 1.63$\times$ and 2.33$\times$ as long as \MANTA at 4 and 6~GB, respectively, and 1.10$\times$ and 2.24$\times$ as long as unfiltered \ARMS.
On \prtwitter at 6~GB, filtered \ARMS takes only 1.01$\times$ as long as \MANTA and 1.03$\times$ as long as unfiltered \ARMS.
Together, these ablations show that adding the filter alone does not reproduce \MANTA's gains, and that virtual-step features improve the learned policy on both workloads.

%% file: figures/6_2_c220g5_eval.tex
\begin{figure*}[t]
\centering

\plotlegend{0.014641\linewidth}{%
  \includegraphics[height=2.5ex]{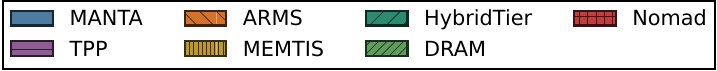}%
}

\hspace*{\fill}%
\begin{subfigure}[t]{0.245\textwidth}
	\centering
	\includegraphics[width=\linewidth]{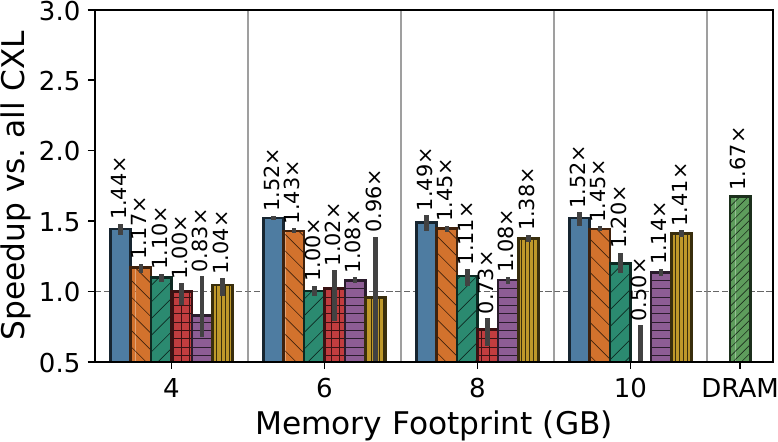}
	\caption{\bckron}
    \label{fig:6_2_c220g5-bc-kron}
\end{subfigure}%
\hfill%
\begin{subfigure}[t]{0.245\textwidth}
	\centering
	\includegraphics[width=\linewidth]{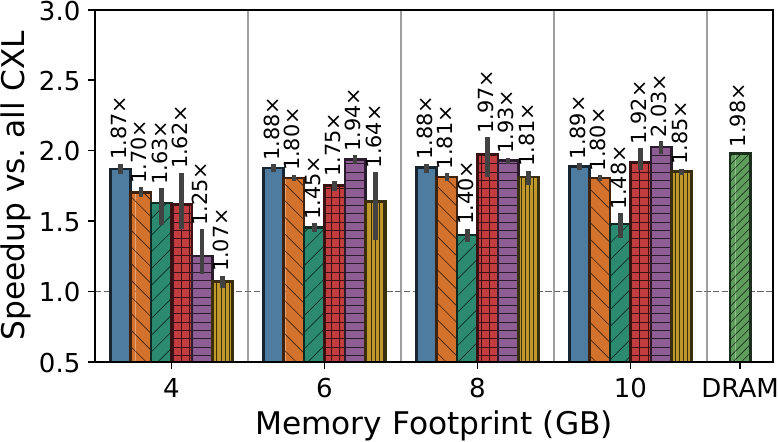}
	\caption{\prkron}
    \label{fig:6_2_c220g5-pr-kron}
\end{subfigure}%
\hfill%
\begin{subfigure}[t]{0.245\textwidth}
	\centering
	\includegraphics[width=\linewidth]{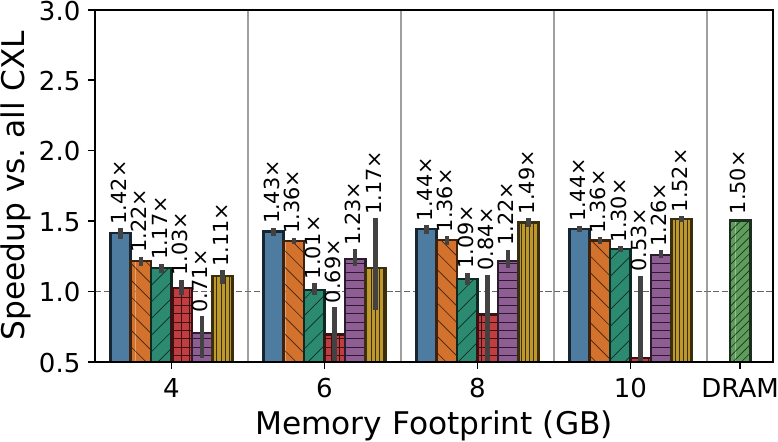}
	\caption{\bctwitter}
    \label{fig:6_2_c220g5-bc-twitter}
\end{subfigure}%
\hfill%
\begin{subfigure}[t]{0.245\textwidth}
	\centering
	\includegraphics[width=\linewidth]{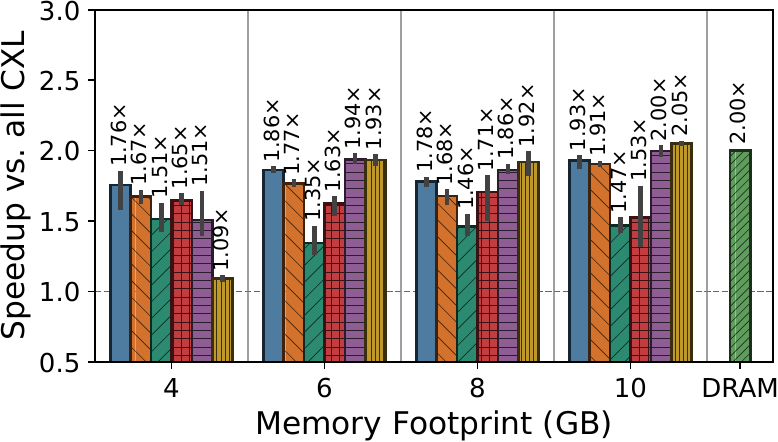}
	\caption{\prtwitter}
    \label{fig:6_2_c220g5-pr-twitter}
\end{subfigure}%
\hspace*{\fill}\par

\hspace*{\fill}%
\begin{subfigure}[t]{0.245\textwidth}
	\centering
	\includegraphics[width=\linewidth]{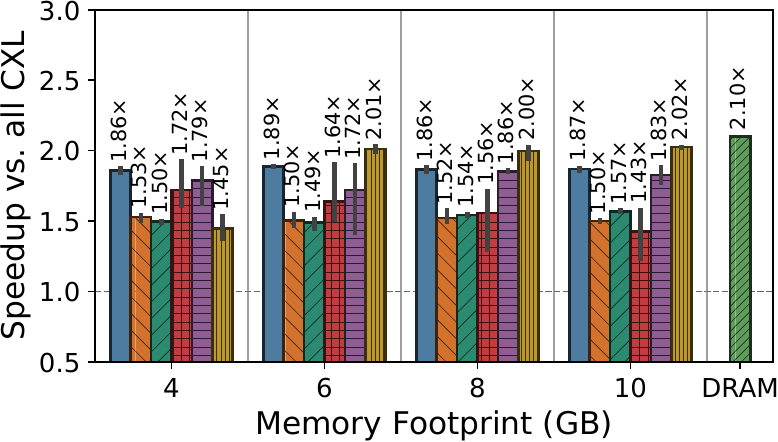}
	\caption{\xsbench}
    \label{fig:6_2_c220g5-xsbench}
\end{subfigure}%
\hfill%
\begin{subfigure}[t]{0.245\textwidth}
	\centering
	\includegraphics[width=\linewidth]{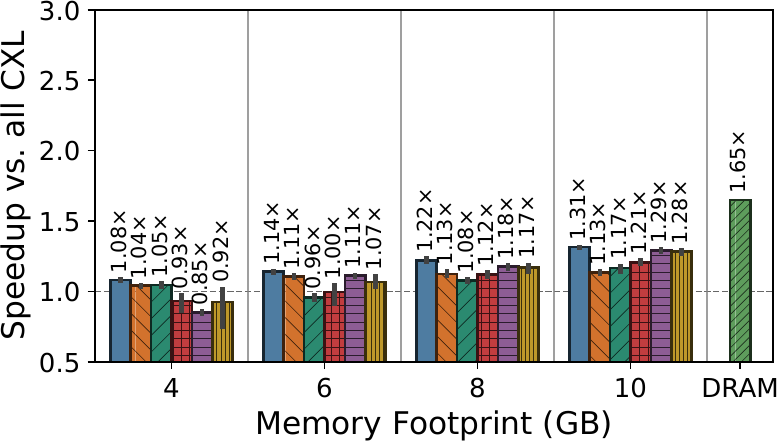}
	\caption{\faiss}
    \label{fig:6_2_c220g5-faiss-10M}
\end{subfigure}%
\hfill%
\begin{subfigure}[t]{0.245\textwidth}
	\centering
	\includegraphics[width=\linewidth]{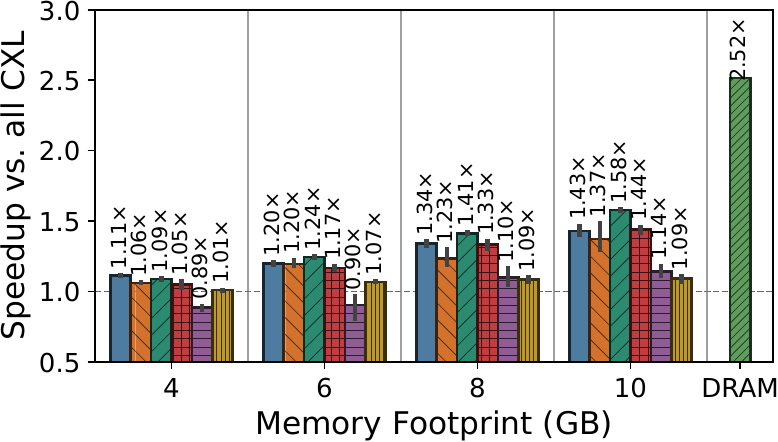}
	\caption{\multigrid}
    \label{fig:6_2_c220g5-mg-d-x}
\end{subfigure}%
\hfill%
\begin{subfigure}[t]{0.245\textwidth}
	\centering
	\includegraphics[width=\linewidth]{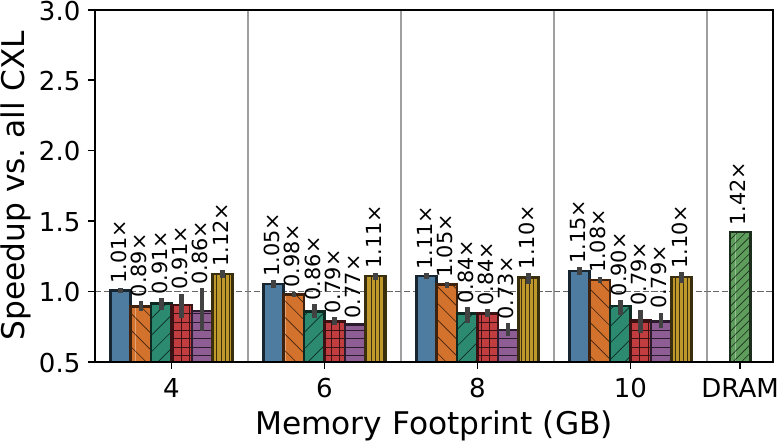}
	\caption{\duckdb}
    \label{fig:6_2_c220g5-duckdb-tpch}
\end{subfigure}%
\hspace*{\fill}\par

\caption{Speedup over all-CXL execution on emulated CXL. Higher is better. \MANTA, \ARMS, and \hybridtier run on Linux 6.2-hybridtier. Other baselines use their patched kernels (Section~\ref{sec:eval:sim-cxl}).}
\Description{Speedup over all-CXL execution (higher is better) on an emulated-CXL \cloudlabnode node with slowed remote NUMA memory serving as the slow tier.
\MANTA, \ARMS, and \hybridtier all use the \texttt{6.2.0-hybridtier+} kernel.
\TPP, \memtis, and \nomad use patched Linux 5.15.0, 5.15.19, and 5.13.0-rc6, respectively.
The panels show speedup for \bckron, \prkron, \bctwitter, \prtwitter, \xsbench, \faiss on OpenAI Embeddings, \multigrid Class D, and \duckdb TPC-H SF100 across memory-capacity settings.}
\label{fig:6_2_c220g5-eval}
\end{figure*}

%% file: figures/6_18_c220g5_eval.tex
\begin{figure*}[t]
\centering

\plotlegend{0.014546\linewidth}{%
  \includegraphics[height=2.5ex]{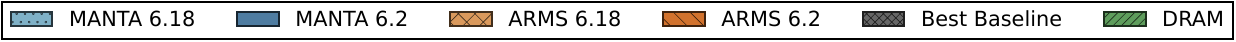}%
}

\hspace*{\fill}%
\begin{subfigure}[t]{0.245\textwidth}
	\centering
	\includegraphics[width=\linewidth]{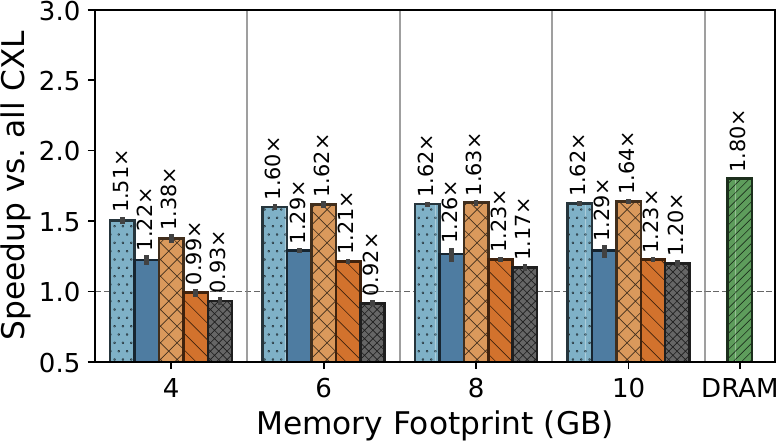}
	\caption{\bckron}
    \label{fig:6_18_c220g5-bc-kron}
\end{subfigure}%
\hfill%
\begin{subfigure}[t]{0.245\textwidth}
	\centering
	\includegraphics[width=\linewidth]{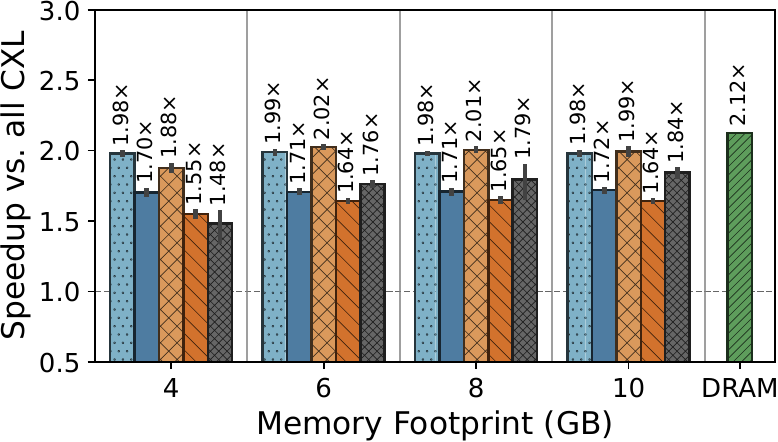}
	\caption{\prkron}
    \label{fig:6_18_c220g5-pr-kron}
\end{subfigure}%
\hfill%
\begin{subfigure}[t]{0.245\textwidth}
	\centering
	\includegraphics[width=\linewidth]{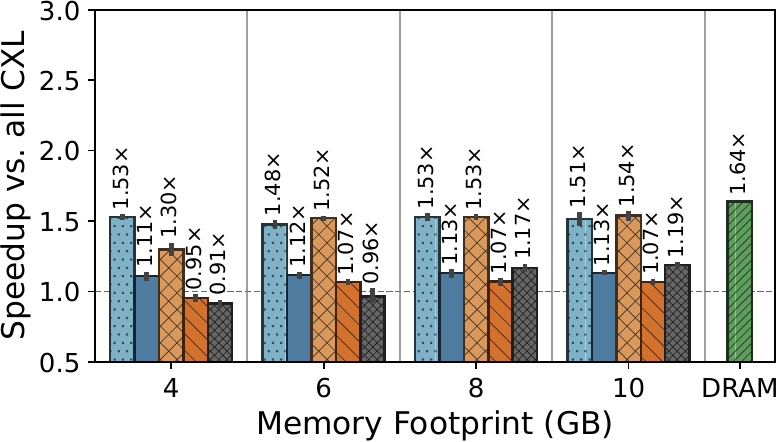}
	\caption{\bctwitter}
    \label{fig:6_18_c220g5-bc-twitter}
\end{subfigure}%
\hfill%
\begin{subfigure}[t]{0.245\textwidth}
	\centering
	\includegraphics[width=\linewidth]{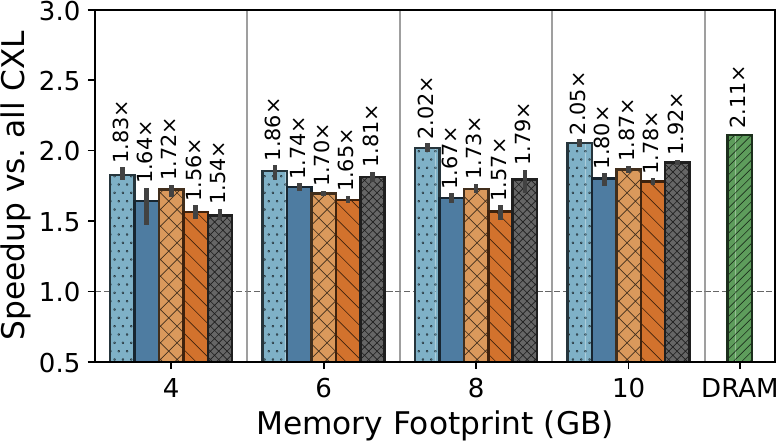}
	\caption{\prtwitter}
    \label{fig:6_18_c220g5-pr-twitter}
\end{subfigure}%
\hspace*{\fill}\par

\hspace*{\fill}%
\begin{subfigure}[t]{0.245\textwidth}
	\centering
	\includegraphics[width=\linewidth]{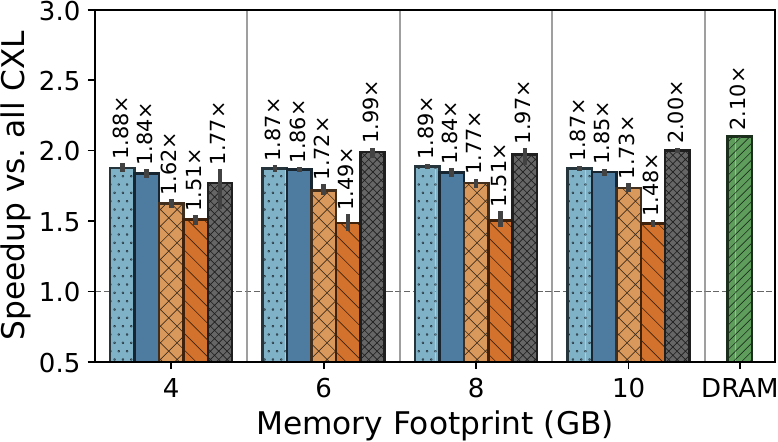}
	\caption{\xsbench}
    \label{fig:6_18_c220g5-xsbench}
\end{subfigure}%
\hfill%
\begin{subfigure}[t]{0.245\textwidth}
	\centering
	\includegraphics[width=\linewidth]{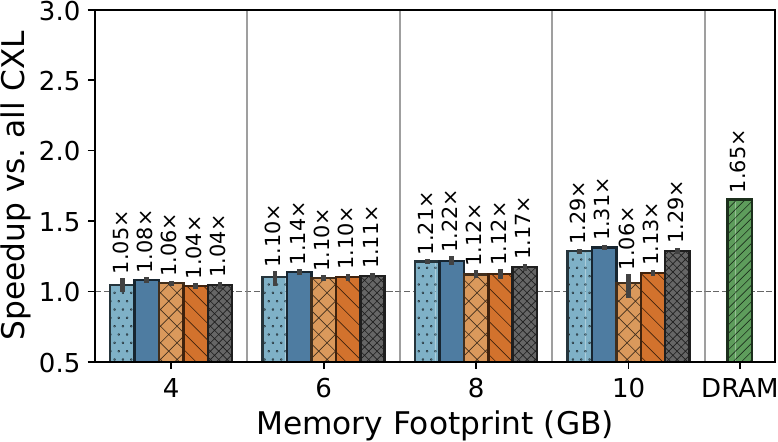}
	\caption{\faiss}
    \label{fig:6_18_c220g5-faiss-10M}
\end{subfigure}%
\hfill%
\begin{subfigure}[t]{0.245\textwidth}
	\centering
	\includegraphics[width=\linewidth]{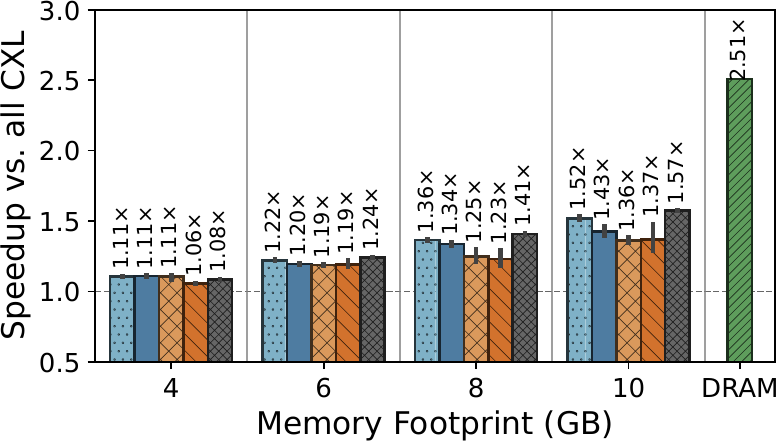}
	\caption{\multigrid}
    \label{fig:6_18_c220g5-mg-d-x}
\end{subfigure}%
\hfill%
\begin{subfigure}[t]{0.245\textwidth}
	\centering
	\includegraphics[width=\linewidth]{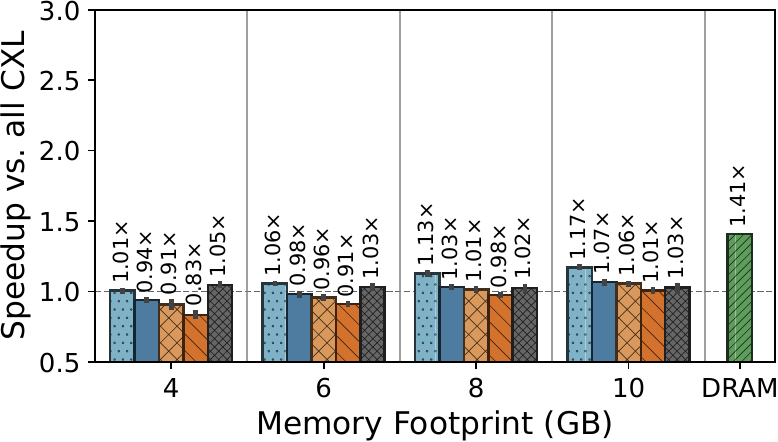}
	\caption{\duckdb}
    \label{fig:6_18_c220g5-duckdb-tpch}
\end{subfigure}%
\hspace*{\fill}\par

\caption{Speedup over all-CXL execution on Linux 6.18. Linux 6.2-hybridtier results, the best baseline, and all-DRAM references are included. Higher is better.}
\Description{Speedup over all-CXL execution (higher is better) on an emulated-CXL \cloudlabnode node with slowed remote NUMA memory serving as the slow tier.
\MANTA and \ARMS run on Linux \texttt{6.18.1-061801-generic}; separately labeled bars show their results on the \texttt{6.2.0-hybridtier+} kernel.
The plots also include a best-baseline comparison and an all-DRAM reference.
The panels show speedup for \bckron, \prkron, \bctwitter, \prtwitter, \xsbench, \faiss on OpenAI Embeddings, \multigrid Class D, and \duckdb TPC-H SF100 across memory-capacity settings.}
\label{fig:6_18_c220g5-eval}
\end{figure*}

%% file: figures/optane_eval.tex
\begin{figure*}[t]
\centering

\plotlegend{0.013343\linewidth}{%
  \includegraphics[height=2.5ex]{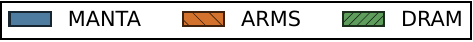}%
}

\hspace*{\fill}%
\begin{subfigure}[t]{0.16\textwidth}
	\centering
	\includegraphics[width=0.9\linewidth]{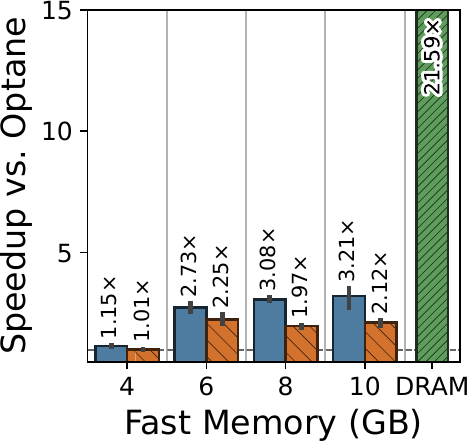}
	\caption{\bckron}
    \label{fig:optane_c220g5-bc-kron}
\end{subfigure}%
\hfill%
\begin{subfigure}[t]{0.16\textwidth}
	\centering
	\includegraphics[width=0.9\linewidth]{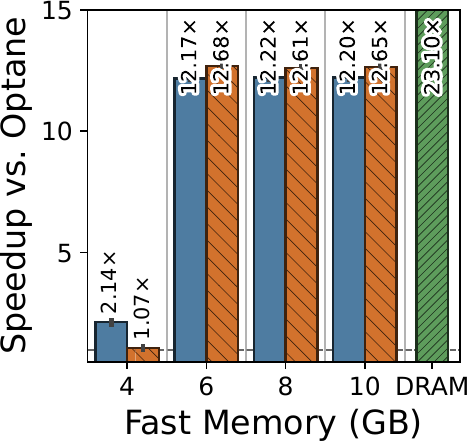}
	\caption{\prkron}
    \label{fig:optane_c220g5-pr-kron}
\end{subfigure}%
\hfill%
\begin{subfigure}[t]{0.16\textwidth}
	\centering
	\includegraphics[width=0.9\linewidth]{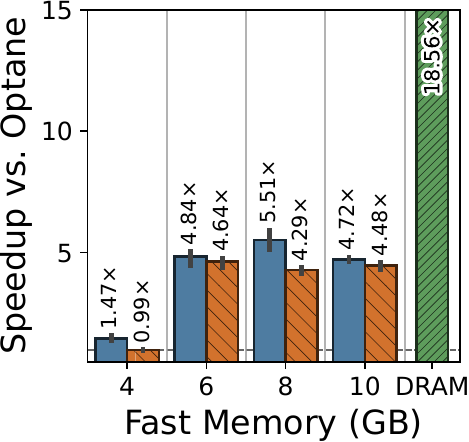}
	\caption{\bctwitter}
    \label{fig:optane_c220g5-bc-twitter}
\end{subfigure}%
\hfill%
\begin{subfigure}[t]{0.16\textwidth}
	\centering
	\includegraphics[width=0.9\linewidth]{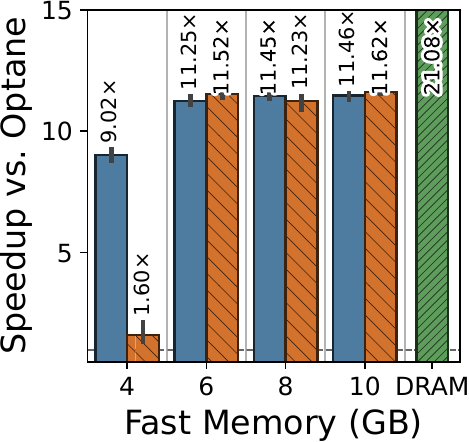}
	\caption{\prtwitter}
    \label{fig:optane_c220g5-pr-twitter}
\end{subfigure}%
\hfill%
\begin{subfigure}[t]{0.16\textwidth}
	\centering
	\includegraphics[width=0.9\linewidth]{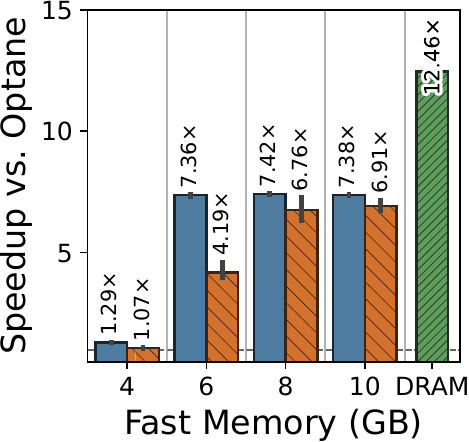}
	\caption{\xsbench}
    \label{fig:optane_c220g5-xsbench}
\end{subfigure}%
\hfill%
\begin{subfigure}[t]{0.16\textwidth}
	\centering
	\includegraphics[width=0.9\linewidth]{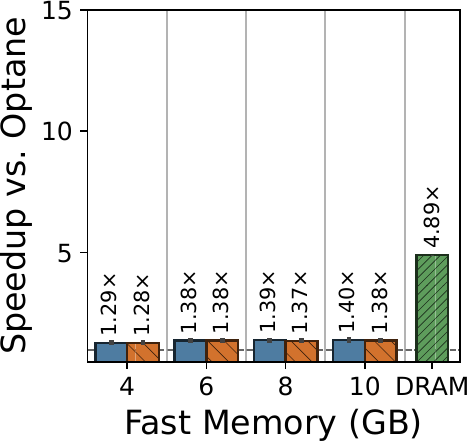}
	\caption{\faiss}
    \label{fig:optane_c220g5-faiss-10M}
\end{subfigure}%
\hspace*{\fill}\par

\caption{Speedup over all-Optane execution, with an all-DRAM reference. Higher is better.}
\Description{Speedup over all-Optane execution for \MANTA and \ARMS on a real Optane persistent-memory tier.
The panels show speedup for \bckron, \prkron, \bctwitter, \prtwitter, \xsbench, and \faiss on OpenAI Embeddings across memory-capacity settings. Higher is better; an all-DRAM reference is included.}
\label{fig:optane_eval}
\end{figure*}

%% file: figures/long_eval.tex
\begin{figure}[t]
\centering

\plotlegend{0.028630\linewidth}{%
  \includegraphics[height=2.5ex]{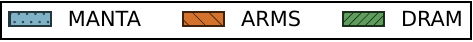}%
}

\hspace*{\fill}%
\begin{subfigure}[t]{0.38\linewidth}
    \centering
    \includegraphics[width=0.9\linewidth]{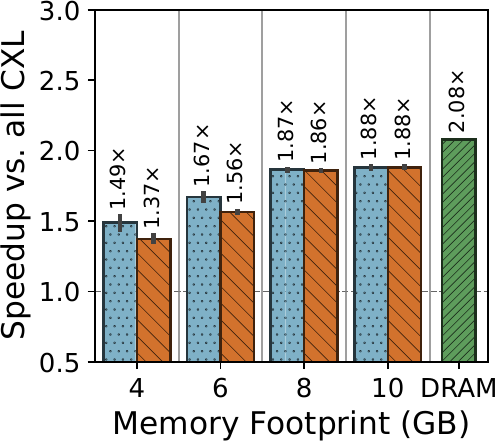}
    \caption{\bctwitter}
    \label{fig:long_eval-bc-twitter}
\end{subfigure}%
\hfill%
\begin{subfigure}[t]{0.38\linewidth}
    \centering
    \includegraphics[width=0.9\linewidth]{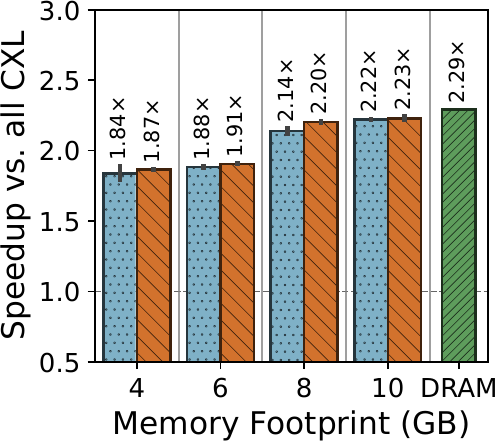}
    \caption{\prtwitter}
    \label{fig:long_eval-pr-twitter}
\end{subfigure}%
\hspace*{\fill}\par

\caption{Longer runs on Linux 6.18. Speedup is over all-CXL execution, with an all-DRAM reference. Higher is better.}
\Description{Two panels compare MANTA with ARMS on longer executions of BC-Twitter and PR-Twitter at four memory capacities on Linux 6.18.
Each panel reports speedup over all-CXL execution and includes an all-DRAM reference.}
\label{fig:long_eval}
\end{figure}

%% file: figures/cross_dataset_eval.tex
\begin{figure*}[t]
\centering

\plotlegend{0.015249\linewidth}{%
  \includegraphics[height=2.5ex]{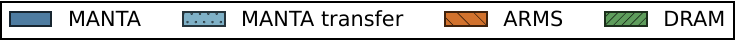}%
}

\hspace*{\fill}%
\begin{subfigure}[t]{0.17\textwidth}
    \centering
    \includegraphics[width=0.9\linewidth]{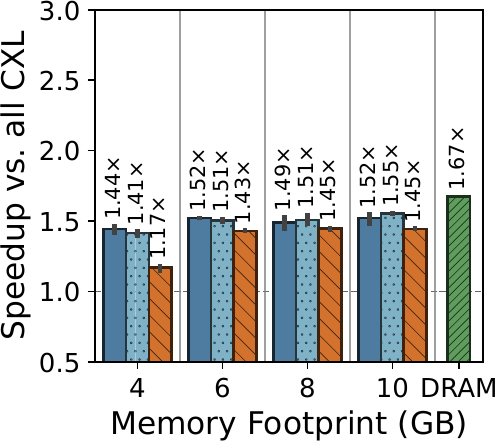}
    \caption{\bckron}
    \label{fig:cross-dataset-eval-bc-kron}
\end{subfigure}%
\hfill%
\begin{subfigure}[t]{0.17\textwidth}
    \centering
    \includegraphics[width=0.9\linewidth]{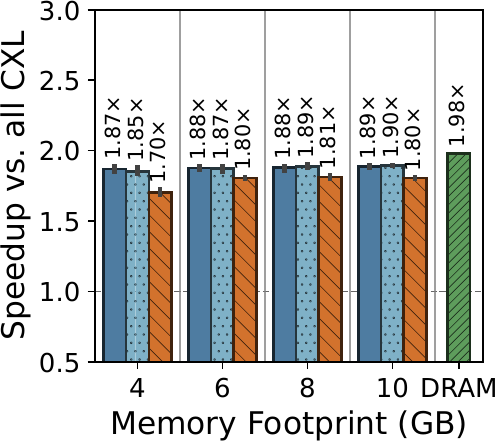}
    \caption{\prkron}
    \label{fig:cross-dataset-eval-pr-kron}
\end{subfigure}%
\hfill%
\begin{subfigure}[t]{0.17\textwidth}
    \centering
    \includegraphics[width=0.9\linewidth]{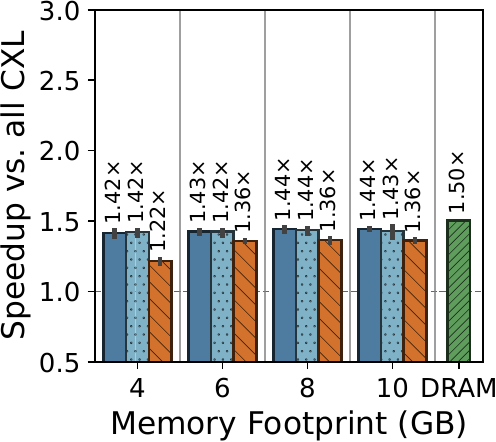}
    \caption{\bctwitter}
    \label{fig:cross-dataset-eval-bc-twitter}
\end{subfigure}%
\hfill%
\begin{subfigure}[t]{0.17\textwidth}
    \centering
    \includegraphics[width=0.9\linewidth]{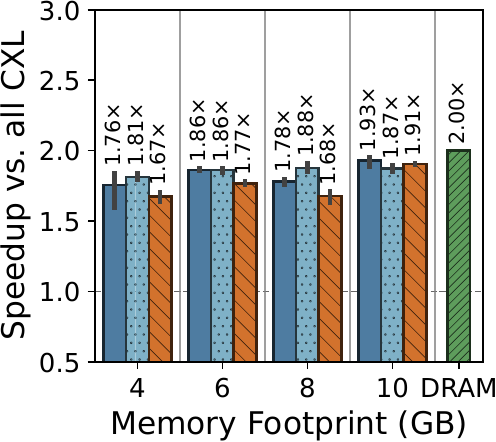}
    \caption{\prtwitter}
    \label{fig:cross-dataset-eval-pr-twitter}
\end{subfigure}%
\hspace*{\fill}\par

\caption{Cross-dataset transfer on Linux 6.2. Panels name the evaluation dataset. Speedup is over all-CXL execution, with an all-DRAM reference. Higher is better.}
\Description{Four panels compare a transferred MANTA model with ARMS on BC-Kron, PR-Kron, BC-Twitter, and PR-Twitter at four memory capacities on Linux 6.2.
Each panel reports speedup over all-CXL execution and includes an all-DRAM reference.}
\label{fig:cross-dataset-eval}
\end{figure*}

%% file: figures/dataset_scale_eval.tex
\begin{figure}[t]
\centering

\plotlegend{0.032171\linewidth}{%
  \includegraphics[height=2.5ex]{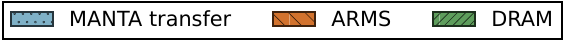}%
}

\hspace*{\fill}%
\begin{subfigure}[t]{0.38\linewidth}
    \centering
    \includegraphics[width=0.9\linewidth]{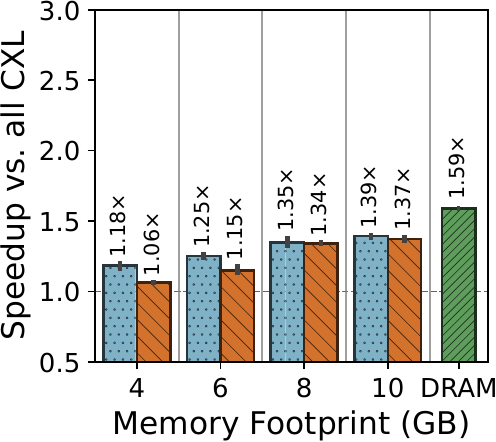}
    \caption{\bckronbig}
    \label{fig:dataset-scale-eval-bc}
\end{subfigure}%
\hfill%
\begin{subfigure}[t]{0.38\linewidth}
    \centering
    \includegraphics[width=0.9\linewidth]{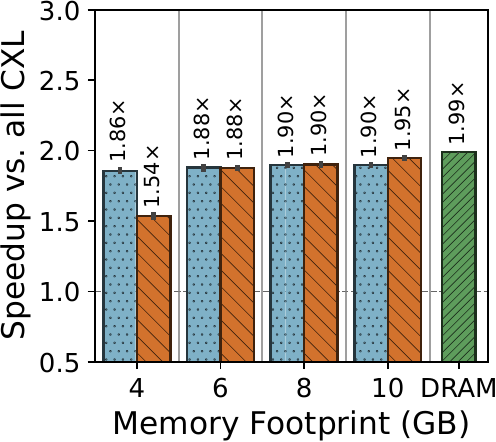}
    \caption{\prkronbig}
    \label{fig:dataset-scale-eval-pr}
\end{subfigure}%
\hspace*{\fill}\par

\caption{Model transfer to an 80~GB Kron graph on Linux 6.18. Speedup is over all-CXL execution. Higher is better.}
\Description{BC and PR speedups over all-CXL execution for transferred MANTA models and ARMS, with an all-DRAM reference.
MANTA's largest gains occur at the smallest fast-tier capacity; PR regresses slightly at the largest.}
\label{fig:dataset-scale-eval}
\end{figure}

%% file: figures/sim_eval.tex
\begin{figure*}[t]
\centering

\plotlegend{0.017638\linewidth}{%
  \includegraphics[height=2.5ex]{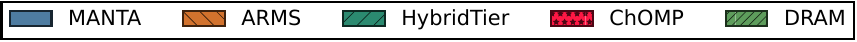}%
}

\hspace*{\fill}%
\begin{subfigure}[t]{0.245\textwidth}
	\centering
	\includegraphics[width=\linewidth]{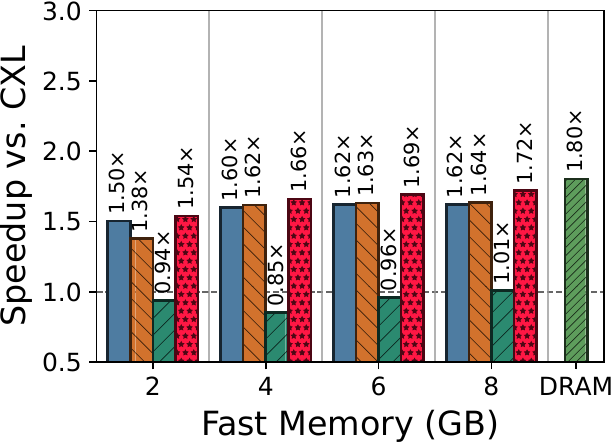}
	\caption{\bckron}
    \label{fig:sim_eval_bc-kron}
\end{subfigure}%
\hfill%
\begin{subfigure}[t]{0.245\textwidth}
	\centering
	\includegraphics[width=\linewidth]{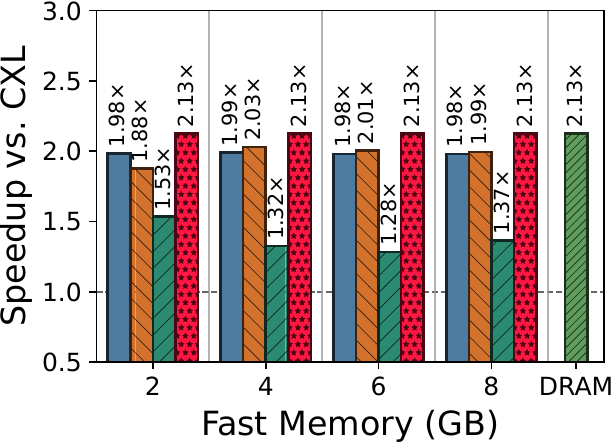}
	\caption{\prkron}
    \label{fig:sim_eval_pr-kron}
\end{subfigure}%
\hfill%
\begin{subfigure}[t]{0.245\textwidth}
	\centering
	\includegraphics[width=\linewidth]{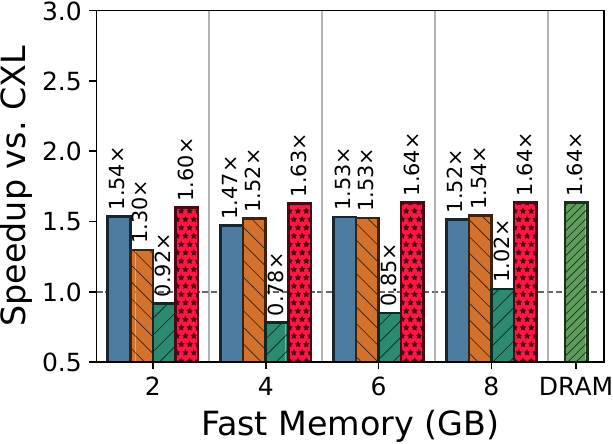}
	\caption{\bctwitter}
    \label{fig:sim_eval_bc-twitter}
\end{subfigure}%
\hfill%
\begin{subfigure}[t]{0.245\textwidth}
	\centering
	\includegraphics[width=\linewidth]{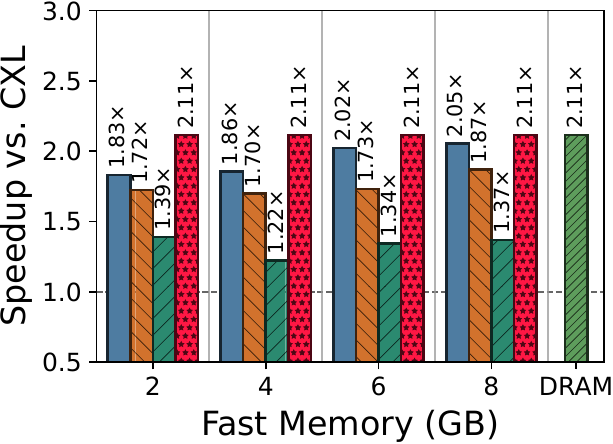}
	\caption{\prtwitter}
    \label{fig:sim_eval_pr-twitter}
\end{subfigure}%
\hspace*{\fill}\par

\hspace*{\fill}%
\begin{subfigure}[t]{0.245\textwidth}
	\centering
	\includegraphics[width=\linewidth]{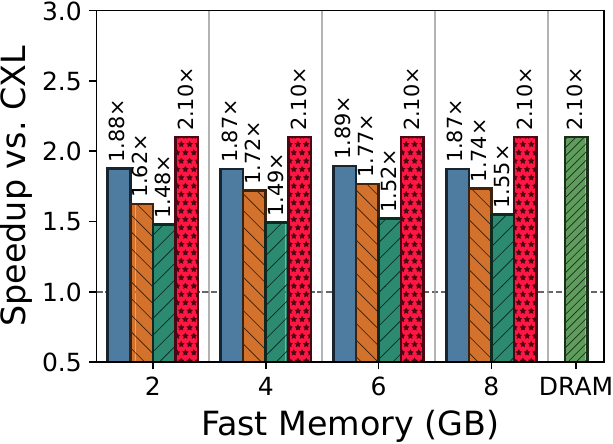}
	\caption{\xsbench}
    \label{fig:sim_eval_xsbench}
\end{subfigure}%
\hfill%
\begin{subfigure}[t]{0.245\textwidth}
	\centering
	\includegraphics[width=\linewidth]{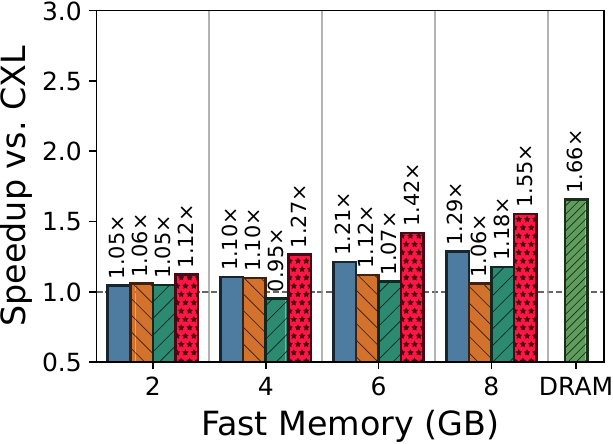}
	\caption{\faiss}
    \label{fig:sim_eval_faiss-10M}
\end{subfigure}%
\hfill%
\begin{subfigure}[t]{0.245\textwidth}
	\centering
	\includegraphics[width=\linewidth]{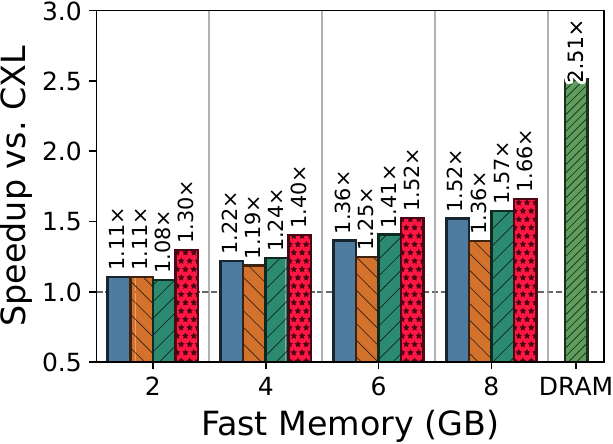}
	\caption{\multigrid}
    \label{fig:sim_eval_mg-d-x}
\end{subfigure}%
\hfill%
\begin{subfigure}[t]{0.245\textwidth}
	\centering
	\includegraphics[width=\linewidth]{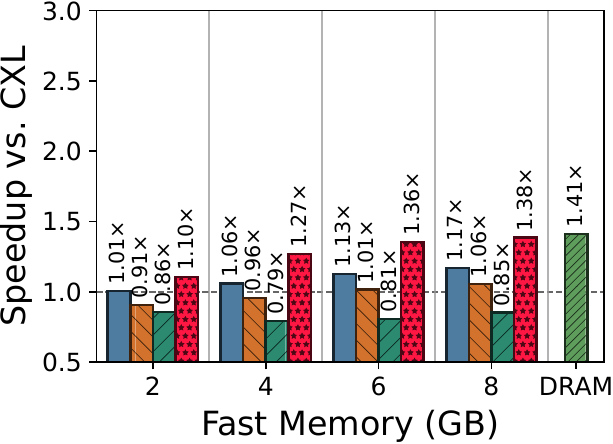}
	\caption{\duckdb}
    \label{fig:sim_eval_duckdb-tpch}
\end{subfigure}%
\hspace*{\fill}\par

\caption{Speedup over simulated all-CXL execution. Higher is better.}
\Description{Eight panels compare measured MANTA, ARMS, and HybridTier runtimes with simulated ChOMP and all-DRAM runtimes on BC Kron, PR Kron, BC Twitter, PR Twitter, XSBench, FAISS, NAS MultiGrid, and DuckDB, normalized to the simulated all-CXL runtime.
Each panel shows DRAM capacities of 2, 4, 6, and 8 GB, plus an all-DRAM reference bar.
ChOMP uses an interval-constrained offline-optimal policy. Higher is better.}
\label{fig:sim_eval}
\end{figure*}

%% file: figures/model_characterization.tex
\begin{figure*}[t]
\centering
{
\footnotesize
\setlength{\tabcolsep}{2pt}
\renewcommand{\arraystretch}{1.0}
\setlength{\aboverulesep}{0.15ex}
\setlength{\belowrulesep}{0.15ex}
\begin{tabular}{@{}llrrrrrrrrrrrrr@{}}
\toprule
\multirow{2}{*}{\textbf{Dataset}} & \multirow{2}{*}{\textbf{Model}} &
\multirow{2}{*}{\shortstack{\textbf{Model build}\\(s) $\downarrow$}} &
\multirow{2}{*}{\shortstack{\textbf{Feature gen.}\\($\mu$s/page) $\downarrow$}} &
\multicolumn{2}{c}{\shortstack{\textbf{Inference}\\($\mu$s/page) $\downarrow$}} &
\multicolumn{2}{c}{\shortstack{\textbf{Total}\\(ms/step) $\downarrow$}} &
\multirow{2}{*}{$R^2$ $\uparrow$} &
\multicolumn{2}{c}{\textbf{Capture@1k} $\uparrow$} &
\multicolumn{2}{c}{\textbf{Recall@1k} $\uparrow$} &
\multicolumn{2}{c}{\textbf{Turnover@1k} $\downarrow$} \\
\cmidrule(lr){5-6}\cmidrule(lr){7-8}\cmidrule(lr){10-11}\cmidrule(lr){12-13}\cmidrule(l){14-15}
 & & & & \textbf{Mean} & \textbf{P95} & \textbf{Mean} & \textbf{P95} & & \textbf{Mean} & \textbf{P5} & \textbf{Mean} & \textbf{P5} & \textbf{Mean} & \textbf{P95} \\
\midrule
\multirow{2}{*}{BC Twitter} & \ARMS & - & \textbf{0.530} & - & - & \textbf{3.613} & \textbf{6.440} & - & 98.66 & 96.90 & 79.85 & 71.10 & 11.98 & 21.86 \\
 & \MANTA & 455 & 1.753 & 1.437 & 1.648 & 22.067 & 25.336 & 0.914 & \textbf{99.07} & \textbf{98.78} & \textbf{81.95} & \textbf{73.16} & \textbf{2.92} & \textbf{5.70} \\
\midrule
\multirow{2}{*}{BC Kron} & \ARMS & - & \textbf{0.460} & - & - & \textbf{5.253} & \textbf{9.416} & - & 96.04 & 82.75 & 86.43 & 63.25 & 5.16 & 24.87 \\
 & \MANTA & 520 & 1.805 & 1.430 & 1.571 & 33.161 & 36.688 & 0.948 & \textbf{98.64} & \textbf{97.37} & \textbf{90.14} & \textbf{78.70} & \textbf{0.93} & \textbf{1.50} \\
\midrule
\multirow{2}{*}{PR Twitter} & \ARMS & - & \textbf{0.439} & - & - & \textbf{3.703} & \textbf{5.472} & - & 99.52 & 99.28 & 44.25 & 36.60 & 28.60 & 33.98 \\
 & \MANTA & 503 & 1.608 & 0.825 & 1.029 & 16.026 & 16.975 & 0.957 & \textbf{99.66} & \textbf{99.33} & \textbf{57.07} & \textbf{43.92} & \textbf{3.06} & \textbf{4.88} \\
\midrule
\multirow{2}{*}{PR Kron} & \ARMS & - & \textbf{0.476} & - & - & \textbf{5.624} & \textbf{8.532} & - & 99.66 & 99.47 & 51.85 & 49.70 & 19.09 & 21.10 \\
 & \MANTA & 726 & 1.649 & 0.791 & 0.948 & 22.970 & 24.117 & 0.966 & \textbf{99.71} & \textbf{99.58} & \textbf{52.30} & \textbf{50.74} & \textbf{5.57} & \textbf{9.60} \\
\midrule
\multirow{2}{*}{DuckDB} & \ARMS & - & \textbf{0.576} & - & - & \textbf{7.409} & \textbf{16.598} & - & 86.83 & 76.37 & 43.49 & 28.00 & 12.89 & 28.90 \\
 & \MANTA & 772 & 1.886 & 1.965 & 2.359 & 34.434 & 51.346 & 0.872 & \textbf{89.69} & \textbf{82.91} & \textbf{43.59} & \textbf{30.06} & \textbf{6.31} & \textbf{13.10} \\
\midrule
\multirow{2}{*}{XSBench} & \ARMS & - & \textbf{0.386} & - & - & \textbf{5.706} & \textbf{11.288} & - & 94.91 & 70.95 & 49.08 & 29.02 & 10.84 & \textbf{19.30} \\
 & \MANTA & 866 & 1.864 & 1.545 & 1.724 & 42.738 & 47.010 & 0.930 & \textbf{99.77} & \textbf{99.74} & \textbf{52.44} & \textbf{50.40} & \textbf{7.00} & \textbf{19.30} \\
\midrule
\multirow{2}{*}{FAISS} & \ARMS & - & \textbf{0.616} & - & - & \textbf{5.289} & \textbf{8.009} & - & 91.49 & 76.87 & \textbf{17.81} & \textbf{14.70} & 25.92 & 27.80 \\
 & \MANTA & 883 & 2.197 & 0.728 & 0.898 & 21.867 & 26.019 & 0.553 & \textbf{98.27} & \textbf{96.34} & 16.52 & 13.70 & \textbf{1.71} & \textbf{4.86} \\
\midrule
\multirow{2}{*}{MG} & \ARMS & - & \textbf{0.692} & - & - & \textbf{13.230} & \textbf{22.911} & - & 97.10 & 95.37 & 3.04 & 1.60 & 30.35 & 55.35 \\
 & \MANTA & 715 & 2.732 & 0.537 & 0.645 & 45.697 & 61.352 & 0.961 & \textbf{99.04} & \textbf{97.31} & \textbf{38.09} & \textbf{6.00} & \textbf{20.56} & \textbf{34.50} \\
\bottomrule
\end{tabular}
}
\captionsetup{skip=\dimexpr 3pt-0.5ex\relax}
\caption{Prediction quality and processing costs. Ranking metrics are percentages. Arrows indicate preferred directions, and bold marks better values.}
\Description{MANTA model build times and \ARMS and \MANTA processing times and prediction quality across eight workloads.}
\label{fig:model-characterization}
\end{figure*}

%% file: figures/6_18_c220g5_cost_ablation_twitter.tex
\begin{figure}[t]
\centering
\plotlegend{0.028556\linewidth}{%
  \includegraphics[width=0.942888\linewidth]{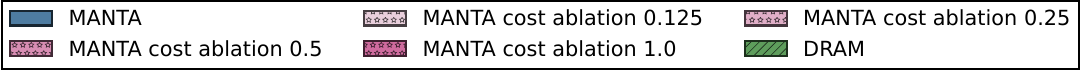}%
}
\hspace*{\fill}%
\begin{subfigure}[t]{0.48\linewidth}
    \centering
    \includegraphics[width=\linewidth]{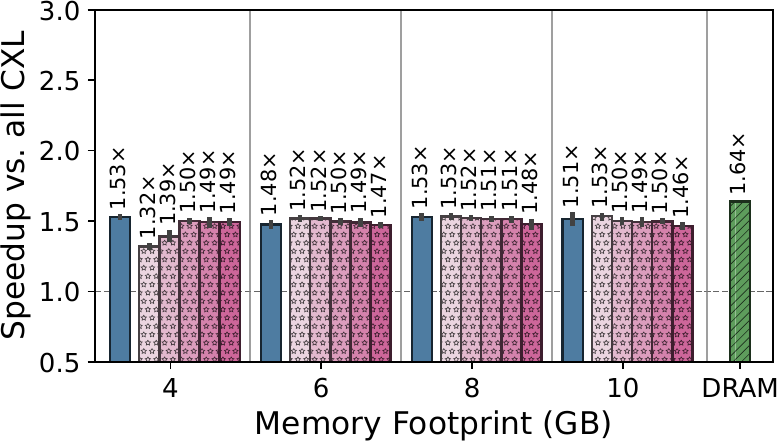}
    \caption{\bctwitter}
    \label{fig:6_18_cost_ablation_c220g5-bc-twitter}
\end{subfigure}%
\hfill%
\begin{subfigure}[t]{0.48\linewidth}
    \centering
    \includegraphics[width=\linewidth]{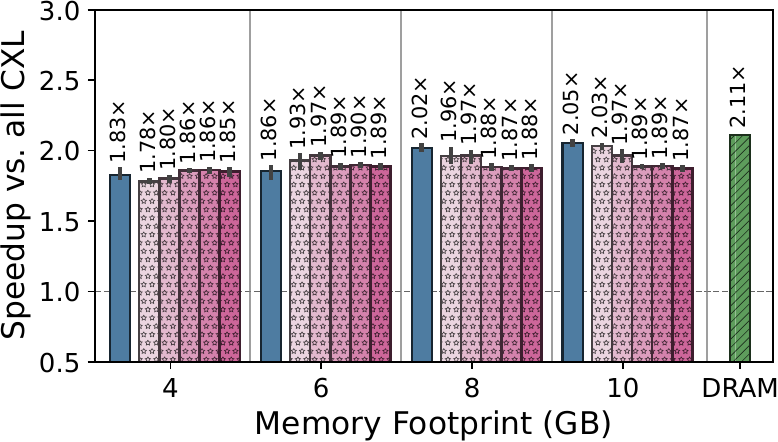}
    \caption{\prtwitter}
    \label{fig:6_18_cost_ablation_c220g5-pr-twitter}
\end{subfigure}%
\hspace*{\fill}\par

\caption{Migration filters on Linux 6.18: default \MANTA ranks versus fixed cost scalers. Speedup is over all-CXL execution. Higher is better.}
\Description{BC Twitter and PR Twitter compare the default MANTA rank filter with fixed migration-cost scalers across fast-tier capacities. Both panels show speedup over all-CXL execution.}
\label{fig:6_18_cost_ablation_c220g5-eval}
\end{figure}

%% file: figures/optane_ablation_eval.tex
\begin{figure}[t]
\centering

\plotlegend{0.033326\linewidth}{%
  \includegraphics[width=0.933348\linewidth]{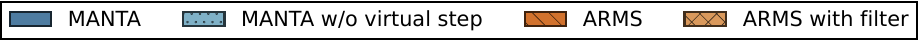}%
}
\hspace*{\fill}%
\begin{subfigure}[t]{.37\linewidth}
	\centering
	\includegraphics[width=0.9\linewidth]{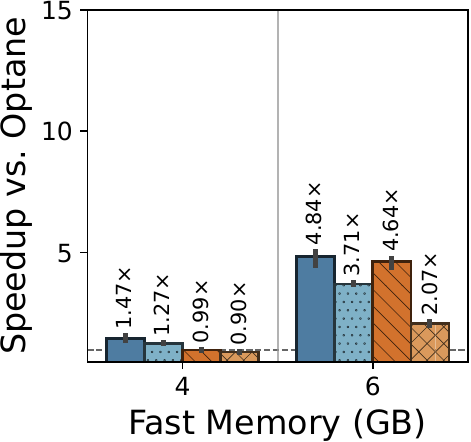}
	\caption{\bctwitter}
    \label{fig:optane_non_model_ablation_bc-twitter}
\end{subfigure}%
\hfill%
\begin{subfigure}[t]{0.37\linewidth}
	\centering
	\includegraphics[width=0.9\linewidth]{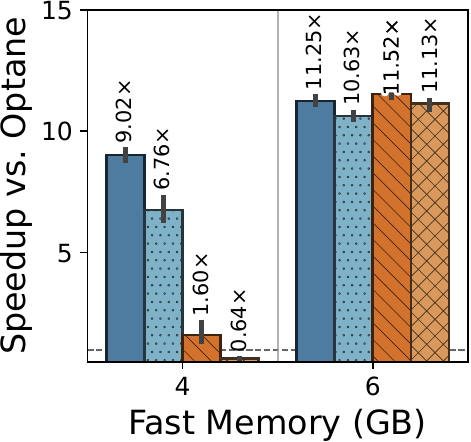}
	\caption{\prtwitter}
    \label{fig:optane_non_model_ablation_pr-twitter}
\end{subfigure}%
\hspace*{\fill}\par

\caption{Optane ablations: replacing virtual steps with 250~ms steps, or adding \MANTA's filter to \ARMS. Speedup is over all-Optane execution. Higher is better.}
\Description{Ablations related to \MANTA's non-learning-based components on the Optane system.
The panels compare \MANTA with non-model variants for \bctwitter and \prtwitter across memory-capacity settings.
The ablations remove \MANTA's virtual step or add \MANTA's filter to \ARMS.
All panels show speedup over all-Optane execution; higher is better.}
\label{fig:optane_non_model_ablation}
\end{figure}

%% file: 70-conclusion.tex
\MANTA predicts future page usefulness to improve online memory tiering.
Our interval-constrained offline optimizer, \chomp, and trace-driven simulator expose opportunities that reactive hotness heuristics miss.
A lightweight learned predictor integrated into \ARMS recovers part of this headroom while retaining practical migration mechanisms.
At 4~GB of fast memory, \MANTA achieves geometric-mean speedups over \ARMS of 1.12$\times$ and 1.08$\times$ on emulated CXL with Linux 6.2 and 6.18, and 1.69$\times$ on Optane.
Its models can transfer to new inputs of the same application, but changes in access behavior require validation and potentially retraining.
The remaining gap to \chomp motivates further work on migration timing and initial placement.

%% file: 71-acknowledgements.tex
We thank our anonymous reviewers and shepherd for their valuable feedback.
This material is based upon work supported by the National Science Foundation under Grant No. 2326576.